\documentclass[prl,twocolumn]{revtex4}
\usepackage{stmaryrd}
\usepackage{marvosym}
\usepackage{url}
\usepackage{subfigure}
\usepackage{amsfonts}
\usepackage{amsmath}
\usepackage{amssymb}
\usepackage{dcolumn}
\usepackage{bm}
\usepackage{epsfig}
\usepackage{graphicx,graphics}
\usepackage{wrapfig}
\usepackage{dcolumn}
\usepackage{fancybox}
\usepackage{euscript}
\usepackage[german,english]{babel}
\usepackage{psfrag}
\usepackage{color}

\begin{document}

\title{Quantum control and entanglement in a chemical compass}
\author{Jianming Cai, Gian Giacomo Guerreschi, and Hans J. Briegel}
\affiliation{$^1$Institut f\"ur Quantenoptik und Quanteninformation
der
\"Osterreichischen Akademie der Wissenschaften, Innsbruck, Austria\\
$^2$Institut f{\"u}r Theoretische Physik, Universit{\"a}t Innsbruck,
Technikerstra{\ss }e 25, A-6020 Innsbruck, Austria}

\begin{abstract}
The radical pair mechanism is one of the two main hypotheses to
explain the navigability of animals in weak magnetic fields,
enabling e.g. birds to \emph{see} the Earth's magnetic field. It
also plays an essential role in the field of spin chemistry. Here,
we show how quantum control can be used to either enhance or reduce
the performance of such a chemical compass, providing a new route to
further study the radical pair mechanism and its applications. We study the role of
quantum entanglement in this mechanism, and demonstrate intriguing
connections between radical-pair entanglement and the magnetic field
sensitivity of the compass. Beyond their immediate application to
the radical pair mechanism, these results also demonstrate how
state-of-the-art quantum technologies could potentially be used to
probe and control biological functions.
\end{abstract}

\maketitle

\textit{Introduction.---} It is known that many species, including
birds, insects and mammals, use the Earth's magnetic field for
orientation and navigation\cite{WJH}. To explain this remarkable
ability, two main hypotheses have been proposed: a
magnetite-based mechanism and a radical pair biochemical reaction mechanism%
\cite{WJH}. Since the radical pair mechanism (RPM) was first
proposed in pioneering work by Schulten \emph{et al.} \cite{Sch78},
a chemical compass model for migratory birds, based on such a
mechanism \cite{Ritz00} has been widely studied. Evidence suggests
that the RPM is indeed linked to the avian magnetoreception
\cite{Ritz04,Wil01}. It was recently demonstrated  in spin chemistry
experiments that a photochemical reaction can act as a compass even
in a magnetic field as weak as the geomagnetic field \cite{Kim08}.
The underlying mechanism in such a chemical compass is clearly of
quantum mechanical nature. However, the detailed role of quantum
interactions, in  giving rise to entanglement and
\mbox{(de-)}coherence, are little understood \cite{Zon}. On the
other hand, one can observe growing interest in the role of quantum
coherence for biological processes in general
\cite{Abb08Hans08Llo09Zur09Gil08}, and specifically in
photosynthesis \cite{Fle0702Fle07Moh08Ple08}. A deeper understanding
of the role of quantum mechanics in biology will eventually come
along with the ability to control biological processes at the level
of individual molecules. In physics, various kinds of quantum
control techniques have been developed, specifically in the field of
quantum information processing and quantum metrology
\cite{Vio99Wal05Uhr09,Sut04Chu05}. The question thus naturally
arises to what extent these or similar techniques could be applied
to test and refine certain biophysical hypotheses, such as the
chemical compass model for animal magnetoreception? Can we use
quantum technologies that have primarily been developed to control
man-made microscopic systems, to study the behavior of living
things --- e.g. birds, fruit flies, or plants --- in a detectable way?

In our work, aiming at the above questions, we will revisit the RPM
and the chemical compass model using concepts and techniques from
quantum information. The RPM can serve both as a magnetometer or as
a compass, depending on the molecular realization. For simplicity,
we will refer to both cases as ``compass'' in the following. First,
we demonstrate that quantum control ideas can be applied to
experiments in spin chemistry and potentially also to study the
magnetoreception of certain animals. We propose several quantum
control protocols that can  be used to either enhance or suppress
the function of a chemical compass. Assuming that the model provides
the correct explanation for magnetoreception of certain species, we
predict that they would loose or regain their orientability in
appropriately designed experiments using such quantum control
protocols -- given that such experiments could be carried out
safely. Our calculations show that the RPM can not only detect weak
magnetic fields, but it is also sensitive to quantum control even
without the presence of a static magnetic field. These results offer
a new means to study experimentally the RPM, also in comparison with
other conceivable mechanisms such as those in man-made magnetometers
\cite{Lukin08,Lukin0802}.

Second, we investigate whether entanglement is a necessary
ingredient in animal magneto-reception, which seems appealing in the
light of the important role this concept has gained in fundamental
discussions on quantum mechanics and its wider implications. As the
sensitivity of the chemical compass depends on the initial state of
the radical pair, it is natural to ask whether it needs to be
quantum mechanically entangled -- thereby excluding any conceivable
classical mechanism -- or whether classical correlations would be
sufficient. We find that the answer largely depends on the radical
pair lifetime. For specific realizations of the RPM, e.g. those in
recent spin-chemistry experiments \cite{Hor07}, entanglement
features prominently and can even serve as a signature of the
underlying spin dynamics. However, when the radical pair lifetime is
extremely long, as it is believed to be the case in the molecular
candidate for magneto reception in European robins
\cite{Ritz09,Sol09}, entanglement does not seem to play a
significant role.

\textit{Radical Pair Mechanism.---} We consider a photochemical
reaction that starts from the light activation of a photoreceptor,
followed by an electron transfer process; two unpaired electrons in
a spin-correlated electronic singlet state are then carried by a
radical pair.
The effective environment of a radical
pair mainly consists of their individual surrounding nuclei.
The Hamiltonian of a radical pair is of the form \cite{Ste89}
\begin{equation}
H=\sum_{k=1,2}H_{k}=-\gamma _{e}\vec{B}\cdot
\sum_{k}\vec{S}_{k}+\sum_{k,j} \vec{S}_{k}\cdot \hat{\lambda}
_{k_{j}}\cdot \vec{I}_{k_{j}}  \label{Hamil}
\end{equation}
where $\gamma _{e}=-g_{e}\mu _{B}$ is the electron gyromagnetic
ratio, $\hat{\lambda}_{k_{j}}$ denote the hyperfine coupling tensors
and $\vec{S}_{k}$, $\vec{I}_{k_{j}}$ are the electron and nuclear
spin operators respectively.

The initial state of a radical pair is assumed to be the singlet
state $|\mathbb{S}\rangle =\frac{1}{\sqrt{2}}(\left\vert \uparrow
\downarrow \right\rangle -\left\vert \downarrow \uparrow
\right\rangle )$, which subsequently suffers from de-coherence
through the hyperfine interactions with the environmental nuclear
spins. The initial state of the nuclear spins at room temperature
can be approximated as $\rho _{b}(0)=\bigotimes_{j}
\mathbb{I}_{j}/d_{j}$, where $d_{j}$ is the dimension of the $j$th
nuclear spin. The charge recombination of the radical pair goes
through different channels, depending on the electron-spin state
(singlet or triplet). In particular, the yield of products formed by
the reaction of singlet radical pairs can be calculated as
\cite{Ste89}
\begin{equation}
\Phi_{s}(t)=\int_{0}^{t}r_{c}(t)f(t)dt  \label{YieldOfProducts}
\end{equation}%
where $r_{c}(t)$ is the radical re-encounter probability
distribution, and $f(t)=\langle \mathbb{S}|\rho
_{s}(t)|\mathbb{S}\rangle $ is the fidelity between the electron
spin state $\rho_{s}(t)$ at time $t$ and the singlet state. The
ultimate activation yield $\Phi_{s} \equiv \Phi_{s}(t\rightarrow
\infty )$ in cryptochrome is believed to affect the visual function
of animals \cite{Ritz00}.

We have followed the established theory for the dynamics of the RPM
\cite{Sch77,Ste89} and computed the full quantum dynamics of the
combined system of electron spins and nuclear spins. Technically, we
employ the Chebyshev polynomial expansion method \cite{Dob03} to
numerically calculate the exact evolution operator
$U_{k}(t)=\exp{(-iH_{k}t)}$, and thereby all relevant physical
quantities. We first consider the well-studied photochemical
reaction of pyrene (Py-$h_{10}$) and N,N-dimethylaniline
(DMA-$h_{11}$) \cite{Sch77}, for which the hyperfine couplings are
isotropic \cite{Hor07}, and the tensor $\hat{\lambda}_{k_{j}}$ in
(\ref{Hamil}) simplifies to a number $\lambda_{k_{j}}$. We study
the role of entanglement in this radical pair reaction and
propose new experiments based on quantum control. We then
generalize our results to the cryptochrome radical pair of
FADH$^\bullet$-O$_{2}^{\bullet -}$, which is the molecular candidate
believed to be involved in avian magnetoreception
\cite{Ritz09,Sol09}. We thereby show that our protocols work also
for anisotropic hyperfine interactions, which are essential for
direction sensitivity of the magnetic field \cite{Ritz00}.

\begin{figure}[tbh]
\begin{center}
\begin{minipage}{9cm}
\hspace{-0.5cm}
\includegraphics[width=4.4cm]{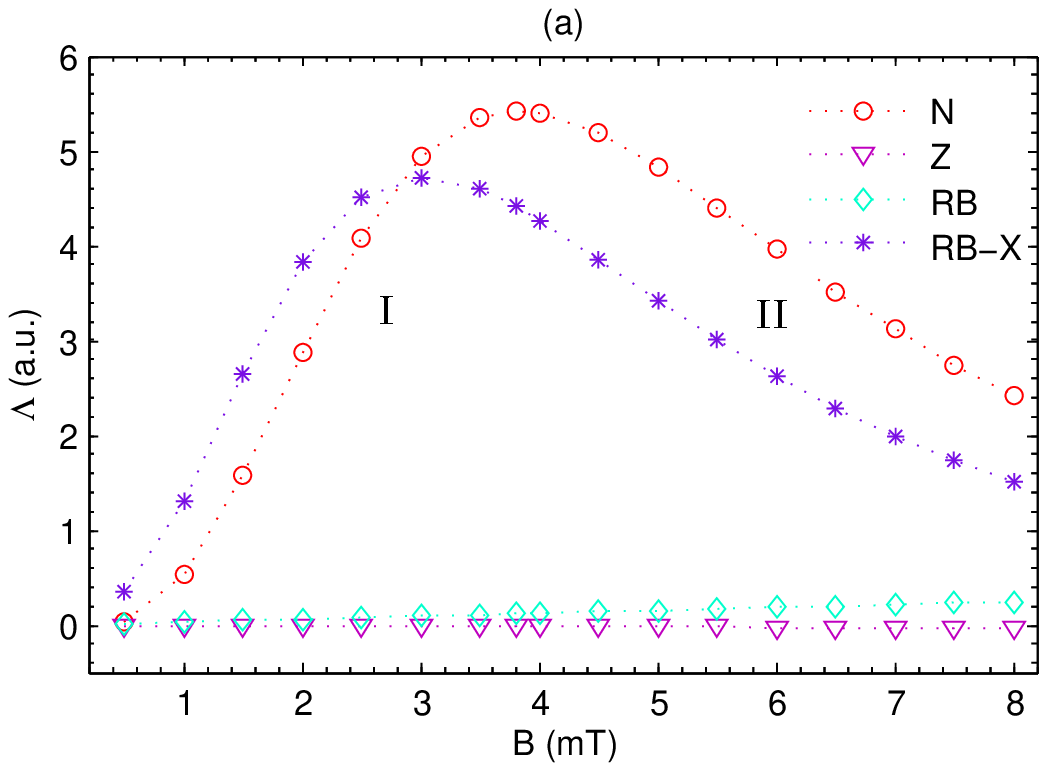}
\hspace{0.3cm}
\includegraphics[width=4.4cm]{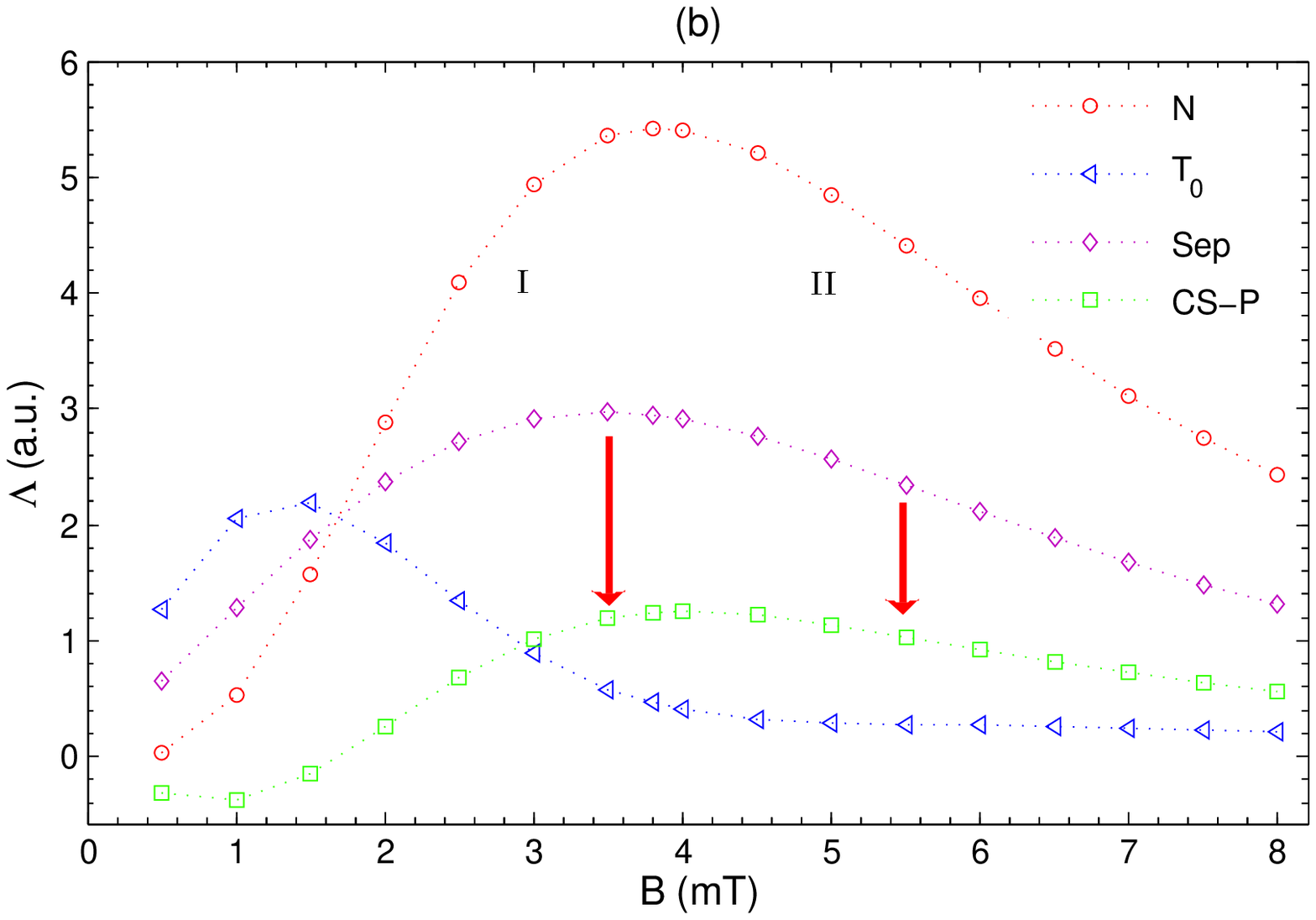}
\end{minipage}
\end{center}
\caption{Magnetic field sensitivity $\Lambda$ of the radical pair
reaction [Py-$h_{10}^{\cdot -}$ DMA-$h_{11}^{\cdot +}$] as a
function of the magnetic field $B$. (a) N: Singlet initial state; Z:
under $Z$ control; RB (RB-X): alternating magnetic field without
(with) $X$ control. (b) N: Singlet initial state; T$_{0}$: Triplet
initial state $|\mathbb{T}_{0}\rangle$; Sep: Optimal sensitivity for
separable initial states; CS-P: applying a $\frac{\protect\pi}{2}$-X
pulse on the initial separable state $\rho_{c} =(\left\vert \uparrow
\downarrow \right\rangle \left\langle \uparrow \downarrow
\right\vert +\left\vert \downarrow \uparrow \right\rangle
\left\langle \downarrow \uparrow \right\vert )/2$. The recombination
rate constant is $k=5.8\times 10^8$ s$^{-1}$ \protect\cite{Hor07},
and the control time is $\protect\tau_{c}=0.5$ ns.} \label{mfs}
\end{figure}

\textit{Magnetic Field Sensitivity under Quantum Control.---} The
magnetic-field sensitivity $\Lambda$ of the radical pair reaction
[Py-$h_{10}^{\cdot -}$ DMA-$h_{11}^{\cdot +}$] is quantified by the
derivative of the activation yield with respect to the magnetic
field strength $B$ \cite{Hor07}, i.e.
\begin{equation}
\Lambda (B)=\frac{\partial \Phi_{s}}{\partial B}\, .
\end{equation}
We assume that the external magnetic field points in the $\hat{z}$
direction. The key ingredient in the RPM are the hyperfine
interactions, which induce a singlet-triplet inter-conversion (mixing)
depending on the magnetic field \cite{Ste89}. Using an exponential
model $r_{c}(t)=k e^{-k t}$ as an example for the re-encounter
probability distribution \cite{Ste89}, we plot in Fig.~\ref{mfs}(a)
the magnetic-field sensitivity $\Lambda$ as a function of $B$. Our
numerical simulation agrees well with the experimental results in
\cite{Hor07}.

Studying the performance of the radical-pair mechanism under quantum
control would allow us to test the role of entanglement and further
details of the RPM in spin chemistry experiments. As a simple
example, consider a periodical pulse sequence with $\pi$-pulses
applied at times $t=m \tau_{c}$ along the $\hat{z}$ direction;
the effective Hamiltonian to the first order is given by
$\bar{H}^{(1)}_{Z}=-\gamma_{e}B\sum_{k}S_{z}^{(k)}+\sum_{k,j}
\lambda_{k_{j}}S_{z}^{(k)}I_{z}^{(k_{j})}$. Such kind of control can
actually enhance the performance of quantum-coherence based
magnetometers, see e.g.\ \cite{Lukin08,Lukin0802}. However, in case
of the RPM, the magnetic-field sensitivity becomes greatly
suppressed, as can be seen in Fig.~\ref{mfs}(a). We can show that,
whenever one applies more general decoupling protocols to promote
quantum coherence in a radical pair reaction, its magnetic-field
sensitivity will generally be reduced \cite{SI}. This demonstrates
that it is in fact the \emph{decay} of coherence, i.e.
\emph{de-}coherence, rather than coherence itself, that plays an
essential role for the magnetic-field detection in RPM, different
from the situation in magnetometers using e.g. NV-centers in diamond
\cite{Lukin0802,Lukin08}.

To demonstrate a potentially positive effect of quantum control on a
chemical compass, we consider a situation where the magnetic field
alternates its direction periodically at times $ t=m \tau_{a}$ which
will disturb the proper functioning of compass. (This situation is
reminiscent of an experiment with birds in an oscillating field
\cite{Ritz04}, even though the cause of the compass disfunction is
here different.) If we now apply $\pi$-X pulses at the same times
$t=m \tau_{a}$, the chemical compass will recover its function as
the transitions between $|\mathbb{S} \rangle$ and
$|\mathbb{T}_{\pm}\rangle$ induced by the residual $xx$ hyperfine
interactions are still affected by the magnetic field [see
Fig.~\ref{mfs}(a)].

\textit{Entanglement and Magnetic Field Sensitivity.---} We have
hitherto assumed, as is usually done, that the radical pair starts
in a perfect singlet state, i.e. that quantum coherence is fully
maintained during the pair creation.
In reality, the initial state $\rho_{s}(0)$ of radical pairs will
never be a perfect singlet (i.e.\ pure state), but a mixed state
with a certain singlet fidelity $f(0)=\langle
\mathbb{S}|\rho_{s}(0)|\mathbb{S}\rangle < 1$. It is known that the
value of $f(0)$ has to be sufficiently close to unity, otherwise the
state may also be described by classical correlations. We therefore
ask: Is entanglement, as a genuine quantum signature, needed at all
to account for the efficiency of the magnetic compass? Or could the
latter be explained by mere classical correlations? To answer this
question, we have randomly chosen $5000$ different initial states
from the set of separable states and calculated the maximal
achievable magnetic field sensitivity for every value of $B$, see
Fig.~\ref{mfs}(b) ($\diamond$). We find that, in the operating
region of the compass (around $B=4$mT), the maximal achievable
sensitivity for separable states stays significantly below the
sensitivity for the singlet state, and this maximum sensitivity is
in fact attained by the classical mixture $\rho_{c} =(\left\vert
\uparrow \downarrow \right\rangle \left\langle \uparrow \downarrow
\right\vert +\left\vert \downarrow \uparrow \right\rangle
\left\langle \downarrow \uparrow \right\vert )/2$. In turn, if
nature is allowed to optimize the initial state from the full set of
states, including the entangled states (e.g. $|\mathbb{S}\rangle$
and $|\mathbb{T}_{0}\rangle$, see Fig.~\ref{mfs}(b)), the optimum
magnetic-field sensitivity will typically be much higher than for
any separable state. On these grounds one can say that entanglement
is indeed helpful, and it is specifically entanglement rather than
mere quantum coherence.

To test experimentally whether the initial state of the radical pair
is indeed (close to) a singlet state, i.e. whether de-coherence can
be neglected during the radical pair creation, one could apply a
$\frac{\pi}{2}$-X pulse as the reaction starts. For an initial
singlet state, which remains invariant under such pulse, the
magnetic field sensitivity will remain unchanged, whereas for an
initial classical mixture it will collapse, see Fig.~\ref{mfs}(b).

As entanglement seemingly plays a role in the RPM with
Py-DMA, we have studied its dynamics and its quantitative connection to the magnetic-field sensitivity.
Similar to the activation yield, we define $\Phi_{E}=
\int_{0}^{\infty} r_{c}(t) E(t) dt$ to quantify the effective amount
of entanglement that is present in the active radical pairs during
the reaction, where $E(t)$ is chosen to be the entanglement measure
of concurrence \cite{Woo98} at time $t$. The first derivative with
respect to the magnetic field, $\Lambda_{E}=\partial
\Phi_{E}/\partial B$, quantifies how sensitive this effective
entanglement is with respect to variations of the magnetic field.

\begin{figure}[tbh]
\begin{center}
\begin{minipage}{9cm}
\hspace{-0.5cm}
\includegraphics[width=4.4cm]{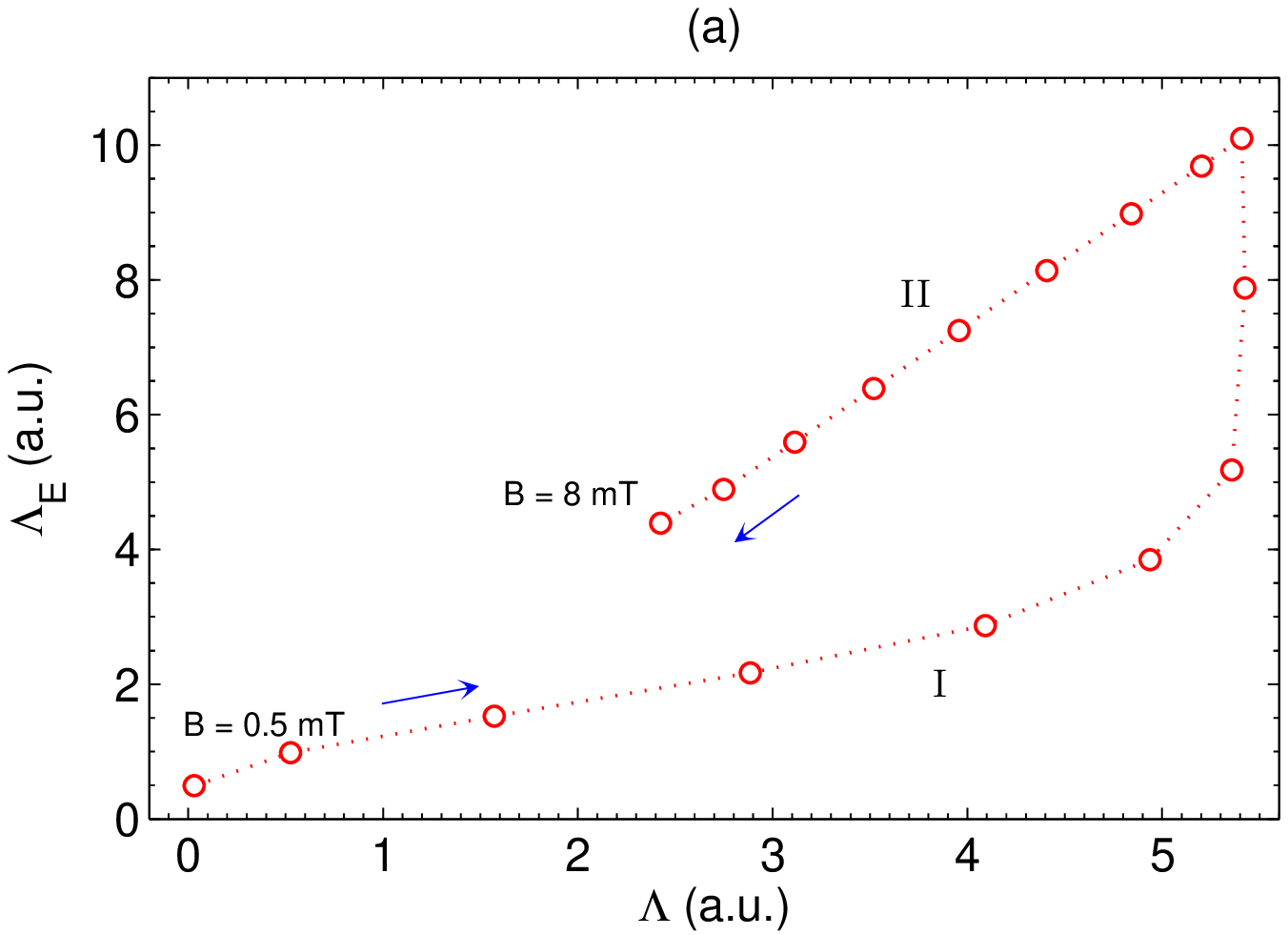}
\hspace{0.3cm}
\includegraphics[width=4.2cm]{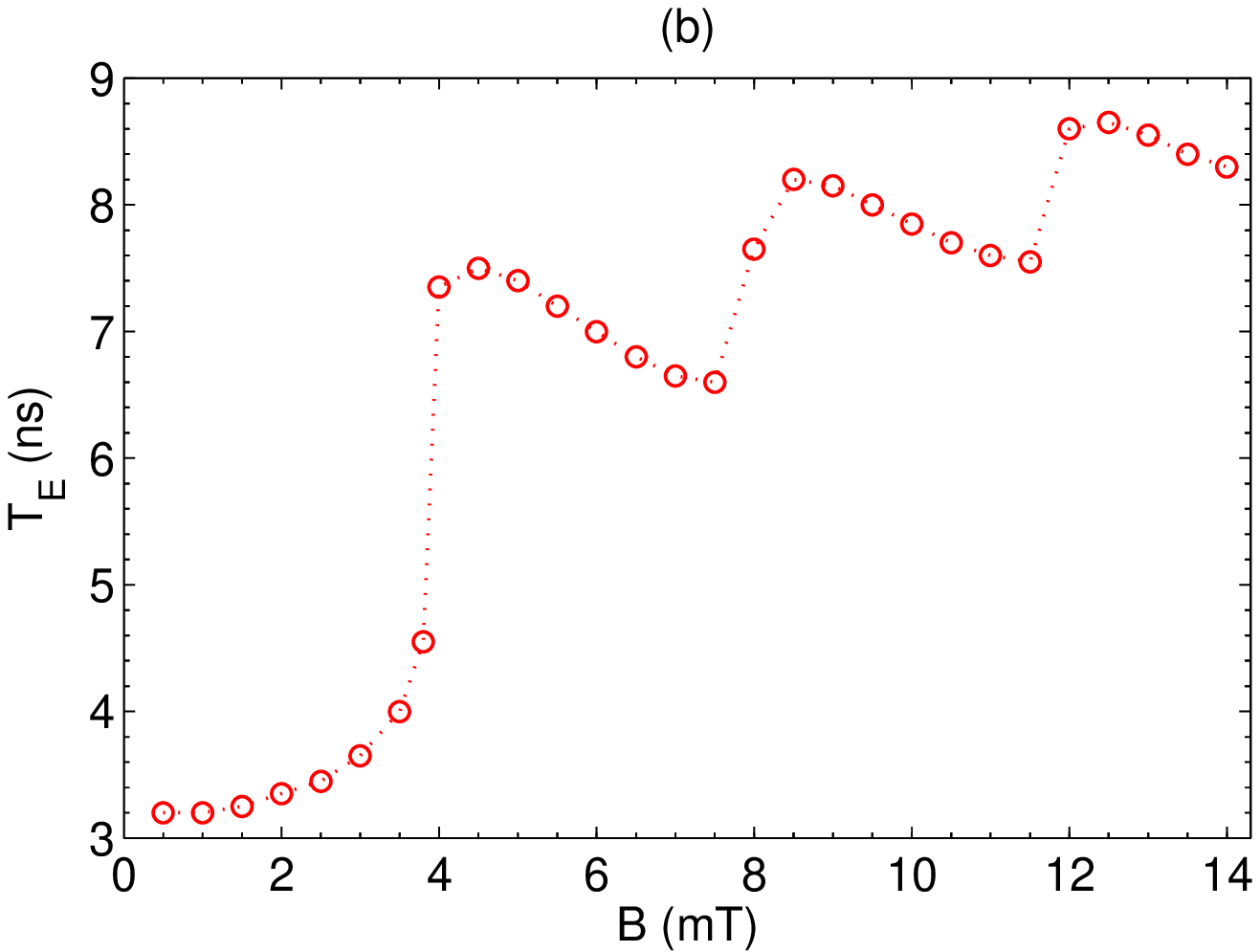}
\end{minipage}
\end{center}
\caption{Connection between quantum entanglement and magnetic field
sensitivity in the radical pair reaction [Py- $h_{10}^{\cdot -}$
DMA-$h_{11}^{\cdot +}$]. (a) Sensitivity of effective entanglement
$\Lambda_{E}$ \textit{vs.} sensitivity of singlet yield $\Lambda$.
The recombination rate constant is $k=5.8\times 10^8 s^{-1}$
\protect\cite{Hor07}. The blue arrows indicate variation of $
\Lambda_{E}$ and $\Lambda$ when the magnetic field changes from
$B=0.5$ mT to $B=8$ mT. (b) Discontinuity of the lifetime of
entanglement $\mathrm{T}_{E}$ as a function of $B$.} \label{entprod}
\end{figure}

In Fig.~\ref{entprod}(a), we see that $\Lambda_{E}$ and $\Lambda$
are correlated in the regions of I and II, displaying
monotonic relations with different linear ratios. This result is
remarkable insofar as that the time during which entanglement exists
is significantly shorter than the reaction time $\mathrm{T}_{r}$ for the
value of $ \Lambda(B,t)=\partial \Phi(t)/\partial B$ to saturate
\cite{SI}. However, it can also be seen from
Fig.~\ref{entprod}(a) that $\Lambda_{E}$ changes dramatically at the
crossover between regions I and II. This step-like behavior relates
to the discontinuity of the entanglement lifetime
$\mathrm{T}_{E}=\max \{t|E(t)> 0\}$ as the magnetic field increases,
see Fig.~\ref{entprod}(b). In the region of I, $\mathrm{T}_{E}$ is
much shorter than the reaction time $\mathrm{T}_{r}$, while it jumps
to a larger value comparable with $\mathrm{T}_{r}$ during the
crossover from the region of I to II. When we further increase the
magnetic field, $\mathrm{T}_{E}$ exhibits more kinks but with less
increment. This effect originates from the finite size of the nuclear spin
bath \cite{Stamp00} of the electron spins, and is a clear signature of the system-environment dynamics underlying the RPM \cite{SI}.

\textit{Applications to Animal Magnetoreception.---} In order to
account for a direction sensitivity of the singlet yield, which is
necessary for compass function, the hyperfine couplings must be
anisotropic \cite{Ritz00}. Here we consider an example of such a
radical pair, FADH$^\bullet$-O$_{2}^{\bullet -}$, which was proposed
as a likely molecular candidate underlying the magnetoreception of
European robins \cite{Ritz09,Sol09}, but it may also play this role
in other species.

\begin{figure}[tbh]
\begin{center}
\begin{minipage}{9cm}
\hspace{-0.5cm}
\includegraphics[width=4.4cm]{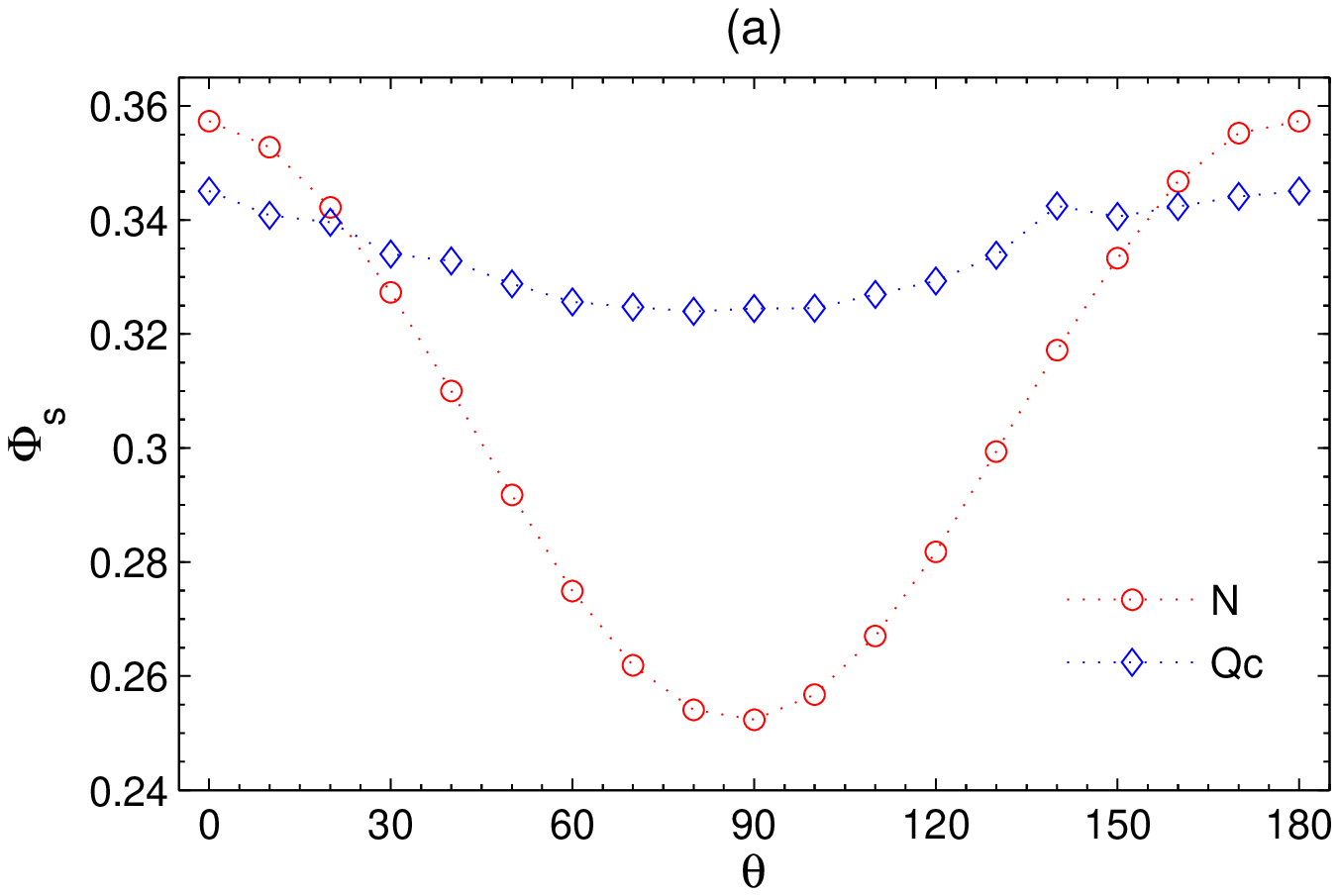}
\hspace{0.0cm}
\includegraphics[width=4.4cm]{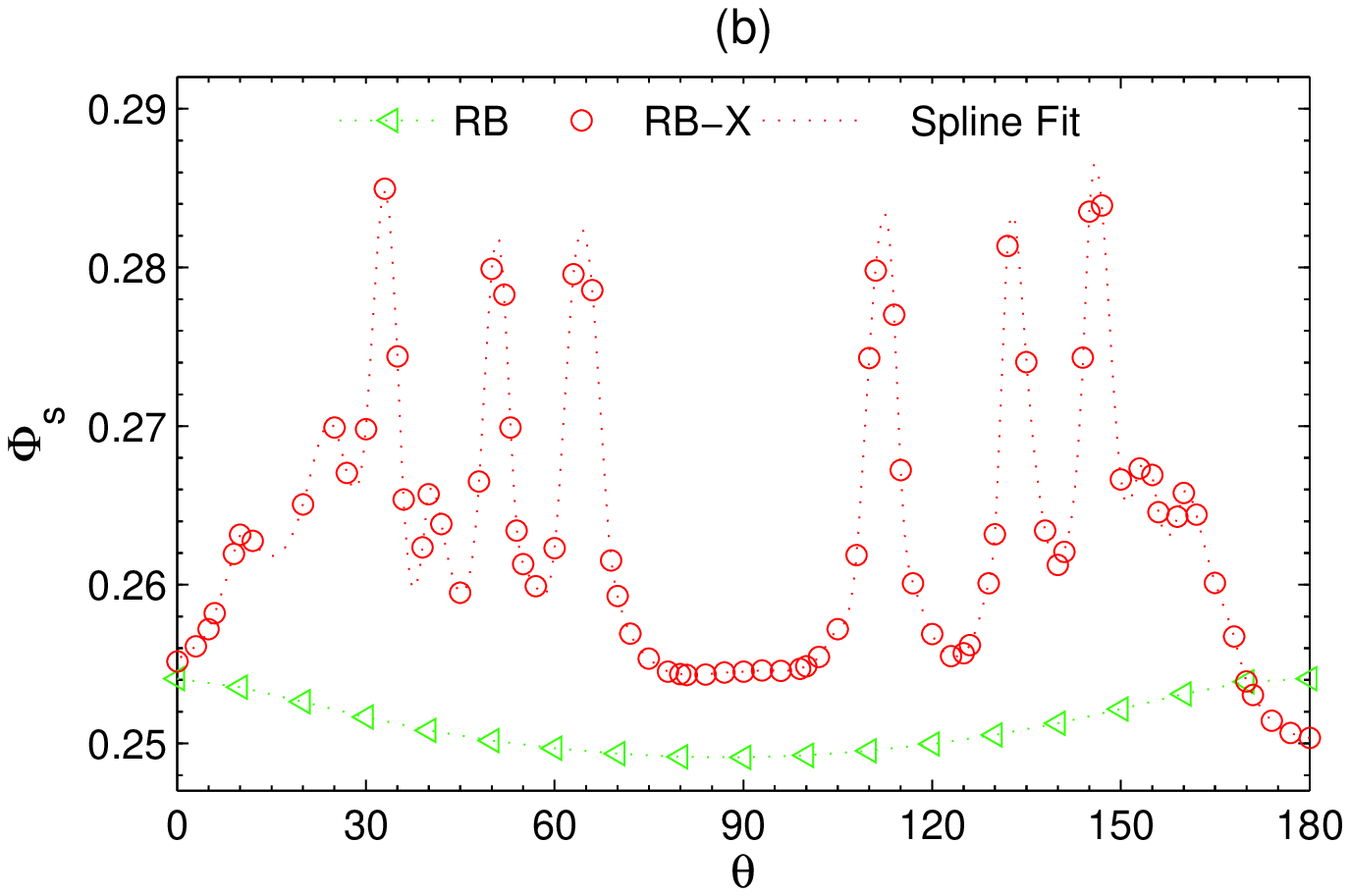}
\end{minipage}
\end{center}
\caption{Singlet yield $\Phi_{s}$ of the radical pair reaction
[FADH$^{\bullet}$-O$_{2}^{\bullet -}$] as a function of the angle
$\theta$ with the magnetic field $B=46$ $\mu T$. (a) N: Without
quantum control; Q$_{c}$: Applying $\pi$-pulses along the direction
of the magnetic field. (b) RB (RB-X): Effect of an alternating
magnetic field, without (with) additional quantum control pulses
perpendicular to the direction of the magnetic field. For
comparison, the RB curve has been shifted downwards by $0.1$. The
recombination rate constant is $k=5\times 10^5$ s$^{-1}$ and the
control times are $\tau_{c}=10$ns and $\protect\tau_{a}=10$ns.} \label{fadh1}
\end{figure}

\begin{figure}[tbh]
\begin{center}
\begin{minipage}{6cm}
\includegraphics[width=6cm]{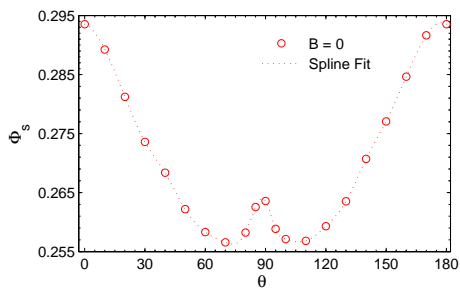}
\end{minipage}
\end{center}
\caption{Singlet yield $\Phi_{s}$ of the radical pair reaction
[FADH$^{\bullet}$-O$_{2}^{\bullet -}$] as a function of the angle
$\theta$, relative to the direction in which the $\pi$-pulses are
applied. There is no external static magnetic field, i.e. $B=0$. The
recombination rate constant is $k=5\times 10^5$ s$^{-1}$, and the
control time is $\protect\tau_{c}=100$ ns.} \label{mqc}
\end{figure}

The direction of the magnetic field in (\ref{Hamil}) with respect to
the reference frame of the immobilized radical pair is described by
two angles $(\theta,\phi)$, i.e. $\vec{B}=B(\sin\theta \cos\phi,$
$\sin\theta \sin\phi, \cos \theta)$. Without loss of the essential
physics, we here assume that $\phi=0$, and investigate the
dependence of the singlet yield $\Phi_{s}$ on the angle $\theta$
when we apply quantum control. First, it can be seen from
Fig.~\ref{fadh1} (a) that the angular dependence of the singlet
yield is much suppressed if one applies $\pi$-pulses along the same
direction as the magnetic field, which can distinguish the RPM from
other potential mechanisms for magnetoreception
\cite{Lukin08,Lukin0802}. Next, we study the scenario that the
magnetic field changes it direction periodically at times $t=n
\tau_{a}$ as in the previous section. As expected, the angular
dependence is again greatly suppressed, as can be seen from the
lower curve in Fig.~\ref{fadh1}(b). However, if one applies
$\pi$-pulses perpendicular to the direction of the magnetic field
this will re-induce an angular dependence, see Fig.~\ref{fadh1}(b).
Furthermore, we find that even without a static magnetic field,
quantum control can induce an angular dependence of the singlet
yield as shown in Fig.~\ref{mqc}. In other words, if one would be
able to design a behavior experiment with animals that use a
chemical compass to sense the magnetic field, in such an environment
they would lose or regain their orientability, depending on the
applied control fields. This would provide further evidence for the
RPM as the underlying mechanism, and it could help to narrow down
the possible candidates of radical pairs in animal magnetoreception.
It is however not clear how such experiments could be carried out
safely.

Different from the example of Py-DMA
we find that here the entanglement only exists in a time range
($\sim 10$ ns) that is much shorter than the expected radical pair
lifetime ($\sim 2-10 \mu$s). We would thus not expect entanglement
to play a significant role in this context. To check this further,
we have computed the achievable sensitivity of the compass for
different initial states and found that a substantial part of all
separable states can account for an angular dependence that is as
high as (or even higher) than for the singlet state \cite{SI}. This
means that -- in contrast to Py-DMA -- the radical-pair entanglement
does not seem to be a necessary ingredient for a chemical compass
based on FADH$^{\bullet}$-O$_{2}^{\bullet -}$.

\textit{Summary and Outlook.---} We have demonstrated how quantum
control can influence radical pair reactions and the function of a
chemical compass. The presented protocols can in principle be
applied to existing spin chemistry experiments, even though the
implementation of coherent spin control \cite{Ber08,Fuc09} in this
context needs to be further developed. They might also provide a
route for future experiments with biological systems (including
animals or plants) that are expected to exploit the RPM; in this
case a much more careful study would be required, in particular
regards the potential side effects of short control pulses on
biological tissue.

We found interesting connections between entanglement and the
magnetic field sensitivity when the radical pair lifetime is not too
long compared to the coherence time. Otherwise, the role of
coherence and entanglement seem to be insignificant. Whether or not
birds or other animals use entanglement for their ability to orient
themselves in the earth magnetic field remains an open question,
whose answer will depend on the specific molecular realization of
their chemical compass.

As a bio-mimetic application of practical relevance, it would be
interesting to explore the possibility of simulating a radical-pair
mechanism in more controllable quantum systems, such as NV centers
in diamond \cite{Lukin08,Lukin0802,Bala08}, to design an ultra-high
fidelity sensor for the detection of weak fields or forces.

\textit{Acknowledgements.---} We are grateful for the support from
the FWF (Lise Meitner Program, SFB FoQuS).


\onecolumngrid \setcounter{figure}{0} \newpage

\section*{Supplementary Material}

This is supporting material for our paper. We derive the completely
positive map for the dynamics of a central spin coupled to its
surrounding nuclear spins with isotropic hyperfine interactions, and
use it to investigate the evolution of entanglement in the radical
pair reaction. To identify the role of entanglement in the radical
pair mechanism, we randomly choose the initial state from the set of
separable states, to compute the optimum magnetic field sensitivity
for these states and compare it with the sensitivity for an inital
singlet state. Further details are provided to clarify the
connections between quantum entanglement and magnetic field
sensitivity. To illustrate the essential role of the (de-phasing)
nuclear spin environment in a chemical compass, we investigate a
hypothetical reference model of bosonic thermal bath and compare it
with the present results.\newline

\begin{figure}[htb]
\begin{center}
\vspace{0cm}
\begin{minipage}[t]{8cm}
\includegraphics[scale=0.15]{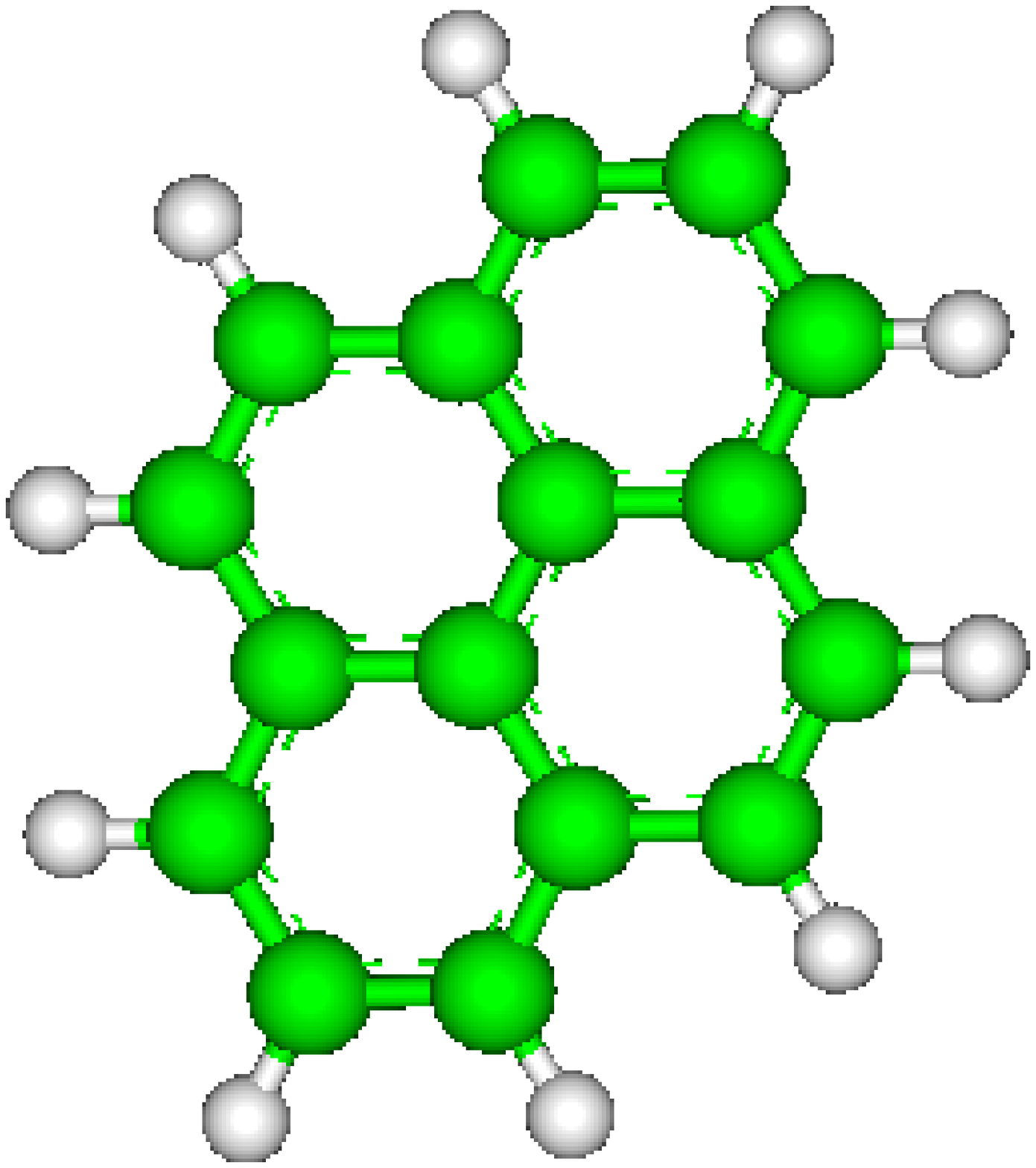}
\hspace{2cm}
\includegraphics[scale=0.15]{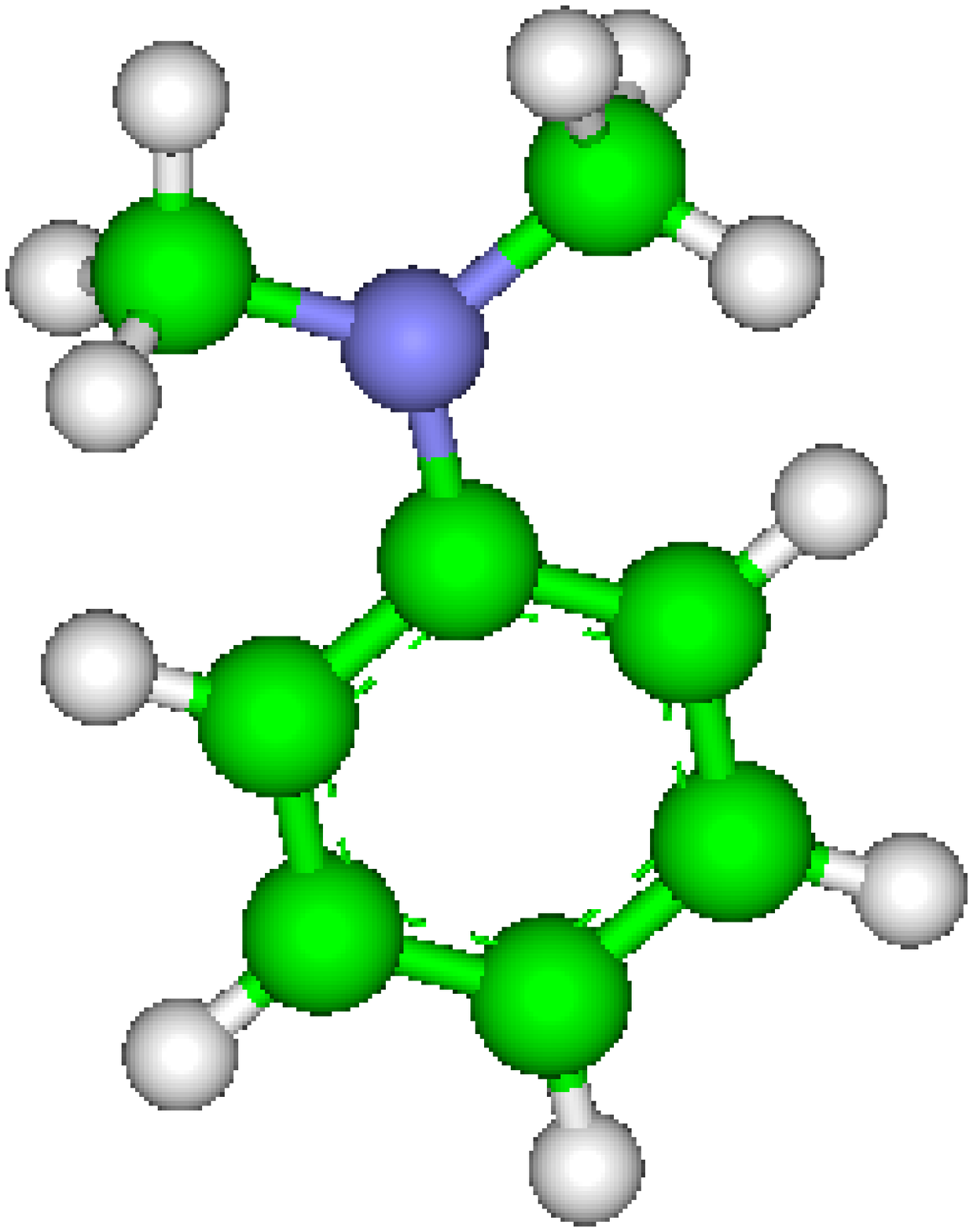}
\end{minipage}
\end{center}
\caption{(Color online) Molecular structures of the radical pyrene (Py-$%
h_{10}$) (left) and N,N-dimethylaniline (DMA-$h_{11}$) (right).
Green: Carbon; Grey: Hydrogen; Blue: Nitrogen.} \label{pdb}
\end{figure}

\textsf{Molecular structures of radical pairs.---} The molecular
structures for the radical pair Py-DMA are displayed in
Fig.~\ref{pdb}, Py-$h_{10}$ has ten spin-$\frac{1}{2} $ hydrogen
nuclei, while the DMA-$h_{11}$ has eleven spin-$\frac{1}{2}$
hydrogen nuclei and one spin-$1$ nitrogen nucleus; the nuclear spin
of carbon is $0$. In our numerical simulations, without loss of
essential features, we have considered three groups of equivalent
nuclei in
each radical that have the largest hyperfine couplings as in \cite{Hor07s}, i.e. the radical Py-$%
h_{10}$ interacts with ten spin-$\frac{1}{2}$ surrounding nuclei
with the hyperfine coupling constants $\lambda _{j_{1}}^{(1)}=0.481$
mT (4$\times $ H), $\lambda_{j_{2}}^{(1)}=0.212$ mT (4$\times $H),
$\lambda
_{j_{3}}^{(1)}=0.103$ mT (2$\times $H) \cite{Hor07s}, and the radical DMA-$%
h_{11}$ is dominantly coupled with seven spin-$\frac{1}{2}$ nuclei
with $\lambda
_{j_{1}}^{(2)}=1.180$ mT (6$\times $H), $\lambda _{j_{2}}^{(2)}=0.520$ mT (1$%
\times $H), and one spin-$1$ nucleus with $\lambda
_{j_{3}}^{(2)}=1.100$ mT (1$\times $ N), see Table 2 of
\cite{Hor07s}.

\noindent The flavin radical FADH$^\bullet$ is displayed in Fig.~\ref%
{fadhstr}. We consider the dominant hyperfine couplings from two
spin-$1$ nitrogen nuclei and three spin-$\frac{1}{2}$ hydrogen
nuclei as in \cite{Hore2003s}.
The superoxide radical O$_{2}^{\bullet -}$ is devoid of the
hyperfine couplings, which is likely to lead to higher sensitivity
against weak magnetic fields \cite{Timmel98s,Ritz09s}.\newline

\begin{figure}[htb]
\vspace{0cm} \includegraphics[width=4cm]{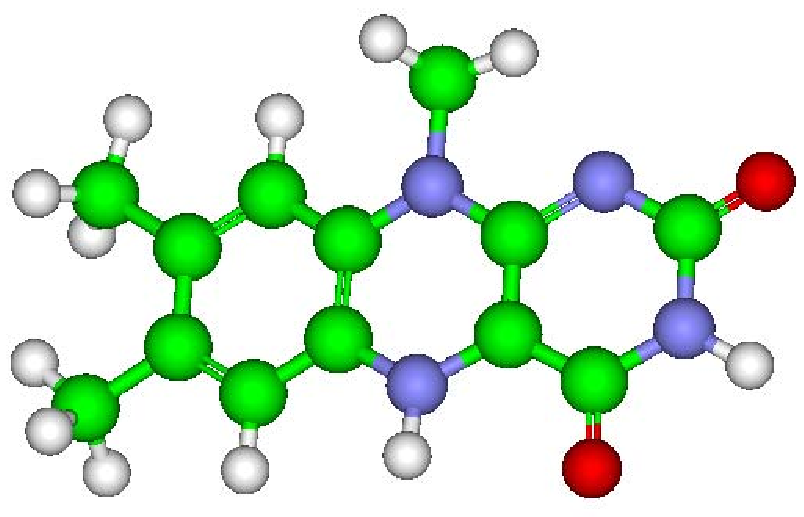}
\caption{(Color online) Molecular structure of the flavin radical FADH$%
^{\bullet}$. Green: Carbon; Grey: Hydrogen; Blue: Nitrogen; Red:
Oxygen.} \label{fadhstr}
\end{figure}

\textsf{Completely positive map for electron spin dynamics.---} We
here derive the completely positive map for the case of isotropic
hyperfine interaction. The Hamiltonian for a central unpaired
electron spin coupled with a nuclear spin bath is written as
\begin{equation}
H_{c}=m_{b}S_{z}+\sum_{k}\lambda _{k}\vec{S}\cdot \vec{I}^{(k)}
\label{Hcent}
\end{equation}
where $m_{b}=-\gamma _{e}B$. The presently available theories for
the central spin problem usually resort to the perturbation
approach, based on certain approximations, e.g. the quasi-static
approximation or the limit of large magnetic fields and/or large
spin bath polarizations. For our present purpose, these
approximations are only of limited use since, in the radical pair
mechanism, one is particularly interested in the region of low
fields, and the number of most relevant surrounding nuclei is $\sim
$ 10, in contrast with $\sim 10^{5} $ in quantum dots.

It is straightforward to show that the total angular momentum of the
electron and nuclear spins, $M_{z}=S_{z}+I_{z}$, where $I_{z}=%
\sum_{k}I_{z}^{(k)}$, is conserved for the Hamiltonian in
Eq.~(\ref{Hcent}), i.e. $[M_{z},H_{c}]=0$. By introducing
$\{|\varphi _{n}^{m}\rangle \}$ as the basis of eigenstates of
$I_{z}$, i.e. $I_{z}|\varphi _{n}^{m}\rangle =n|\varphi
_{n}^{m}\rangle $, where $n$ labels the eigenvalues and $m$ is a
degeneracy index, we can express the initial state of the spin bath
as $\rho
_{b}(0)=\bigotimes_{k}\mathbb{I}_{k}/d_{k}=\frac{1}{d}\sum_{n,m}|\varphi
_{n}^{m}\rangle \langle \varphi _{n}^{m}|$, where $d=\prod_{k}d_{k}$
is the total dimension of all the (relevant) nuclear spins. Thus,
under the coherent evolution $U_{c}=\exp (-itH_{c})$, the joint
state of the central spin and the nuclear spins evolves as
\begin{eqnarray}
\left\vert \uparrow \right\rangle |\varphi _{n}^{m}\rangle
&\rightarrow &\left\vert \uparrow \right\rangle |\varphi
_{mn}^{0}\rangle +\left\vert
\downarrow \right\rangle |\varphi _{mn}^{1}\rangle \\
\left\vert \downarrow \right\rangle |\varphi _{n}^{m}\rangle
&\rightarrow &\left\vert \uparrow \right\rangle |\varphi
_{mn}^{-1}\rangle +\left\vert \downarrow \right\rangle |\varphi
_{mn}^{0^{\prime }}\rangle
\end{eqnarray}%
wherein $\left\vert \downarrow \right\rangle $ and $\left\langle
\downarrow \right\vert $ denote the eigenstates of
$S_{z}=\frac{\hbar}{2}\sigma _{z}$, and $|\varphi _{mn}^{i}\rangle $
belongs to the eigenspace of $I_{z}$ associated to the eigenvalue
$n+i$. The fact that the total angular momentum is conserved results
in orthogonality relations for the nuclear spin states:
\begin{equation}
|\varphi _{mn}^{0}\rangle ,|\varphi _{mn}^{0^{\prime }}\rangle \perp
|\varphi _{mn}^{-1}\rangle \perp |\varphi _{mn}^{1}\rangle
\label{ortho}
\end{equation}%
The inner products of these vectors are zero, as they belong to
orthogonal subspaces (or are null vectors). By recalling the
notation $\frac{1+\sigma _{z}}{2}=\left\vert \uparrow \right\rangle
\left\langle \uparrow \right\vert $, $\frac{1-\sigma
_{z}}{2}=\left\vert \downarrow \right\rangle \left\langle \downarrow
\right\vert $, $\sigma _{+}=\left\vert \uparrow \right\rangle
\left\langle \downarrow \right\vert $, $\sigma _{-}=\left\vert
\downarrow \right\rangle \left\langle \uparrow \right\vert $, we
obtain
\begin{eqnarray}
\mu _{0+} =\mathrm{Tr}\left[ U_{c}\left( \frac{1+\sigma
_{z}}{2}\otimes
\frac{\mathbb{I}}{d}\right) U_{c}^{\dagger }(\sigma _{+}\otimes \mathbb{I})%
\right] \varpropto \mathrm{Tr}\left[ U_{c}\left(
\sum_{n,m}\left\vert \uparrow \right\rangle |\varphi _{n}^{m}\rangle
\langle \varphi _{n}^{m}|\left\langle \uparrow \right\vert \right)
U_{c}^{\dagger }(\sigma _{+}\otimes \mathbb{I})\right]
=\mathrm{Tr}\sum_{n,m}|\varphi _{mn}^{1}\rangle \langle \varphi
_{mn}^{0}| =0
\end{eqnarray}%
in which we have used the relation in Eq.~(\ref{ortho}). In a
similar way, one can show that $\mu _{0-}=\mu _{1\pm }=$ $\mu _{\pm
0}=\mu _{\pm 1}=\mu _{++}=\mu _{--}=0$. Moreover, it is easy to
verify that
\begin{equation*}
\mu _{00}=\mu _{11}=\frac{1}{2}+\frac{1}{4d}Tr\left( U_{c}\sigma
_{z}U_{c}^{\dagger }\sigma _{z}\right)
\end{equation*}%
Thus, the dynamics of the central spin, which is calculated by
tracing out
its spin bath degrees of freedom as $\rho _{s}(t)=\mathrm{Tr}%
_{b}\{e^{-iH_{c}t}[\rho _{s}(0)\otimes \rho _{b}(0)]e^{iH_{c}t}\}$,
can be explicitly expressed as
\begin{eqnarray*}
\xi (t):\quad \left\vert \uparrow \right\rangle \left\langle
\uparrow \right\vert &\rightarrow &a_{t}\left\vert \uparrow
\right\rangle \left\langle \uparrow \right\vert +(1-a_{t})\left\vert
\downarrow
\right\rangle \left\langle \downarrow \right\vert \\
\left\vert \downarrow \right\rangle \left\langle \downarrow
\right\vert &\rightarrow &(1-a_{t})\left\vert \uparrow \right\rangle
\left\langle \uparrow \right\vert +a_{t}\left\vert \downarrow
\right\rangle \left\langle
\downarrow \right\vert \\
\left\vert \uparrow \right\rangle \left\langle \downarrow
\right\vert &\rightarrow &\kappa _{t}\left\vert \uparrow
\right\rangle \left\langle
\downarrow \right\vert \\
\left\vert \downarrow \right\rangle \left\langle \uparrow
\right\vert &\rightarrow &\kappa _{t}^{\ast }\left\vert \downarrow
\right\rangle \left\langle \uparrow \right\vert
\end{eqnarray*}%
from which we can obtain the completely positive map for the central
spin dynamics. If we rewrite the evolution operator
$U_{c}=\exp(-itH_{c})$ in the form of $U_{c}=\sum_{\mu ,\nu }|\mu
\rangle _{c}\langle \nu |\otimes U_{\mu \nu }$, we get the above
dynamic parameters $a_{t}=\mathrm{Tr}(U_{00}U_{00}^{\dag })$, and
$\kappa _{t}= \mathrm{Tr}(U_{00}U_{11}^{\dag })$.\newline

\begin{figure}[tbh]
\begin{center}
\epsfig{file=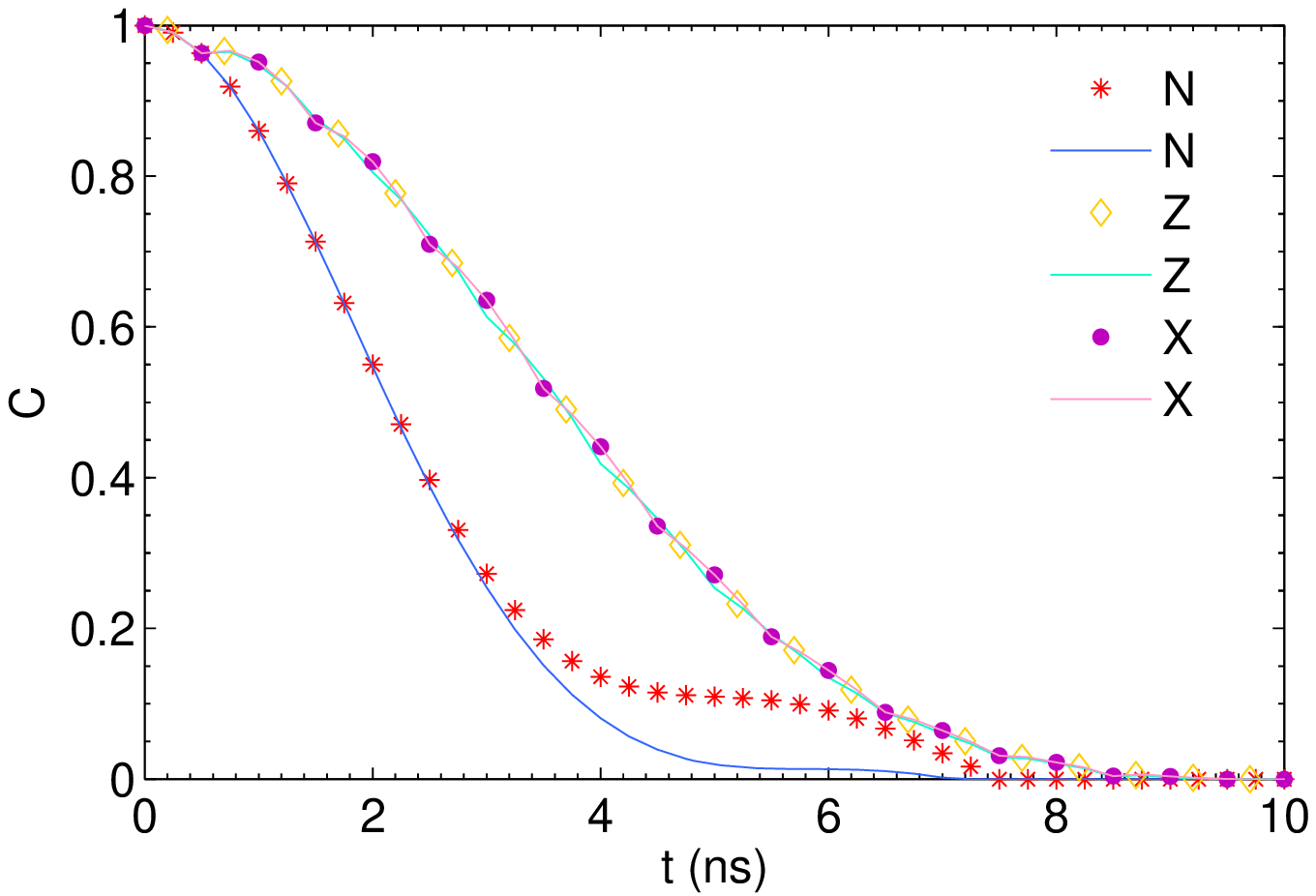,width=8cm}
\end{center}
\caption{Decay of the entanglement in a radical pair reaction [Py-$%
h_{10}^{\cdot -}$ DMA-$h_{11}^{\cdot +}$] under different types of
quantum control. (N) without control; (Z) under $Z$ control, (X)
under $X$ control. The curves are the estimated best lower bounds
from the singlet fidelity, the symbols denote the values from
numerical simulation. The magnetic field is $B=4.5$ mT, and the
control time is $\protect\tau=0.5$ ns.} \label{entdyn}
\end{figure}

\textsf{Dynamics of quantum entanglement in Py-DMA.---} In Fig.~\ref{entdyn}%
, we plot the evolution of entanglement between two unpaired
electron spins in the radical pair reaction [Py-$h_{10}^{\cdot -}$
DMA-$h_{11}^{\cdot +}$]. We have used the concurrence \cite{Woo98s}
as the measure of two-spin entanglement, which vanishes on separable
states and assumes its maximum value, 1, on maximally entangled
states such as the singlet state. We also compare the exact value of
entanglement with the estimated best lower bound of entanglement
$\varepsilon(t)=\inf_{\rho}\{E(\rho)|\mathrm{Tr}(\rho
\mathcal{P}_{s})=f_{s}(t)\}=\max \{0,2 f_{s}(t)-1\}$. The agreement
between them is good (even though not perfect). This fact supports
our statement about how one could possibly estimate the amount of
entanglement from experimentally accessible information.

\noindent We also plot the dynamics of entanglement under X and Z
control in Fig.~\ref{entdyn}. It can be seen that entanglement
survives indeed for a longer time if quantum control is
applied.\newline

\textsf{Protecting coherence is not helpful.---} In the quantum
coherence
based magnetometer, e.g. with NV centers in diamond \cite{Lukin08s,Lukin0802s}%
, the sensitivity is indeed dependent on the coherence time, i.e.
the longer the coherent time is the better the sensitivity. In this
section, we use the simple example of the radical pair with only
isotropic hyperfine couplings to show that this is not the case in
the present model.

As we have described in the main text, for the Z control, we
dynamically decouple the xx and yy hyperfine interactions while
keeping the magnetic-field dependent Zeeman interactions.
Nevertheless, the magnetic field sensitivity is still much
suppressed. This phenomenon can be understood as follows. The
residual hyperfine couplings along the longitudinal direction (i.e.
zz hyperfine couplings) only induce the transitions between the
singlet state $|\mathbb{S}\rangle$ and one specific triplet state $|
\mathbb{T}_{0}\rangle$, while these two eigenstates are degenerate
and their energies are independent of the magnetic field. In this
case, the singlet-triplet interconversion is actually not influenced
by the magnetic field, the effects of which are thus not detectable
through the singlet yield. We will prove, in the following, that a
chemical compass will loose its function, if one uses general
dynamical decoupling protocols to promote the electron spin
coherence.

Assume that, at time $t_{0}$, the electron spins and the surrounding
nuclear spins are in some state $\rho (t_{0})=\rho _{0}$. The
activation yield during a short time interval $[t_{0},t_{0}+\tau ]$
is
\begin{equation}
\Phi (t_{0},\tau )=\int_{t_{0}}^{t_{0}+\tau }r_{c}(t)f_{s}(t)dt
\label{sts}
\end{equation}%
with the singlet fidelity $f(t)=\langle \mathbb{S}|\rho _{s}(t)|\mathbb{S}%
\rangle $. We then write its first derivative with respect to the
magnetic field as
\begin{equation*}
\Lambda (t_{0},\tau )=\frac{\partial \Phi (t_{0},t_{0}+\tau )}{\partial B}%
=\int_{t_{0}}^{t_{0}+\tau }r_{c}(t)\frac{\partial f_{s}(t)}{\partial
B}dt
\end{equation*}%
which obviously determines the ultimate magnetic field sensitivity as $%
\Lambda =\sum_{m}\Lambda (m\tau ,\tau )$ by summing up $\Lambda
(t_{0},\tau ) $ for all time intervals $\left[ t_{0},t_{0}+\tau
\right] =\left[ m\tau ,(m+1)\tau \right] $, $m=0,1,2,...$. The
singlet fidelity at time $t\in \lbrack t_{0},t_{0}+\tau ]$ is
$f_{s}(t)=\mathrm{Tr}\left( e^{-i\Delta tH}\rho _{0}e^{i\Delta
tH}\mathcal{P}_{s}\right) $ with $\Delta t=t-t_{0}$, and $H$ is the
Hamiltonian of the electron spins together with the nuclear spins as
in Eq.~(L1) \cite{Note1}. By using a perturbation expansions for
small $\Delta t$, we have
\begin{equation}
e^{-i\Delta tH}=\mathbb{I}-i\Delta tH-\frac{(\Delta t)^{2}}{2}%
H^{2}+O((\Delta t)^{3})
\end{equation}%
which enables us to express the singlet fidelity as follows,
\begin{eqnarray}
f_{s}(t)=\mathrm{Tr} \left[ \rho _{0}\left( \mathcal{\
P}_{s}+i\Delta t\left[
H,\mathcal{P}_{s}\right] +\frac{(\Delta t)^{2}}{2}\left[ \left[ H,\mathcal{P}%
_{s}\right] ,H\right] \right) \right] +O(\Delta t^{3})
\end{eqnarray}%
where $[A,B]=AB-BA$. Using the properties of the singlet state that $\frac{%
\partial H}{\partial B}\mathcal{P}_{s}=\mathcal{P}_{s}\frac{\partial H}{%
\partial B}=0$, where $\frac{\partial H}{\partial B}=-\gamma
_{e}(S_{z}^{(1)}+S_{z}^{(2)})$, the first derivative of $f_{s}(t)$
can be written as
\begin{eqnarray}
\frac{\partial f_{s}(t)}{\partial B} =-\frac{(\Delta t)^{2}}{2}\mathrm{Tr}%
\left[ \rho _{0}\left( \frac{\partial H}{\partial B}H\mathcal{P}_{s}+%
\mathcal{P}_{s}H\frac{\partial H}{\partial B}\right)
\right]+O((\Delta t)^{3})
\end{eqnarray}%
If $\rho _{0}=\mathcal{P}_{s}\otimes \frac{\mathbb{I}}{d}$, one can
easily
verify that $\mathrm{Tr}\left[ \rho _{0}\left( \frac{\partial H}{\partial B}%
H \mathcal{P}_{s}+\mathcal{P}_{s}H\frac{\partial H}{\partial B}\right) %
\right] =0$. Thus, if a dynamical decoupling protocol is to protect
the electron spin coherence during the reaction, i.e. keep the spin
state close to the singlet state, we can conclude that $\partial
f_{s}(t)/\partial B\simeq O((\Delta t)^{3})$, and $\Lambda
(t_{0},\tau )$ will be of the fourth order in $\tau $, which is an
order smaller than the one from other general states. We remark that
we do not trivially assume that the system dynamics is frozen by
protecting coherence, but the electron spin state does evolve even
if it is kept closer to the singlet state under decoupling
controls.\newline

\begin{figure}[htb]
\begin{center}
\begin{minipage}[t]{15cm}
\includegraphics[width=6.2cm]{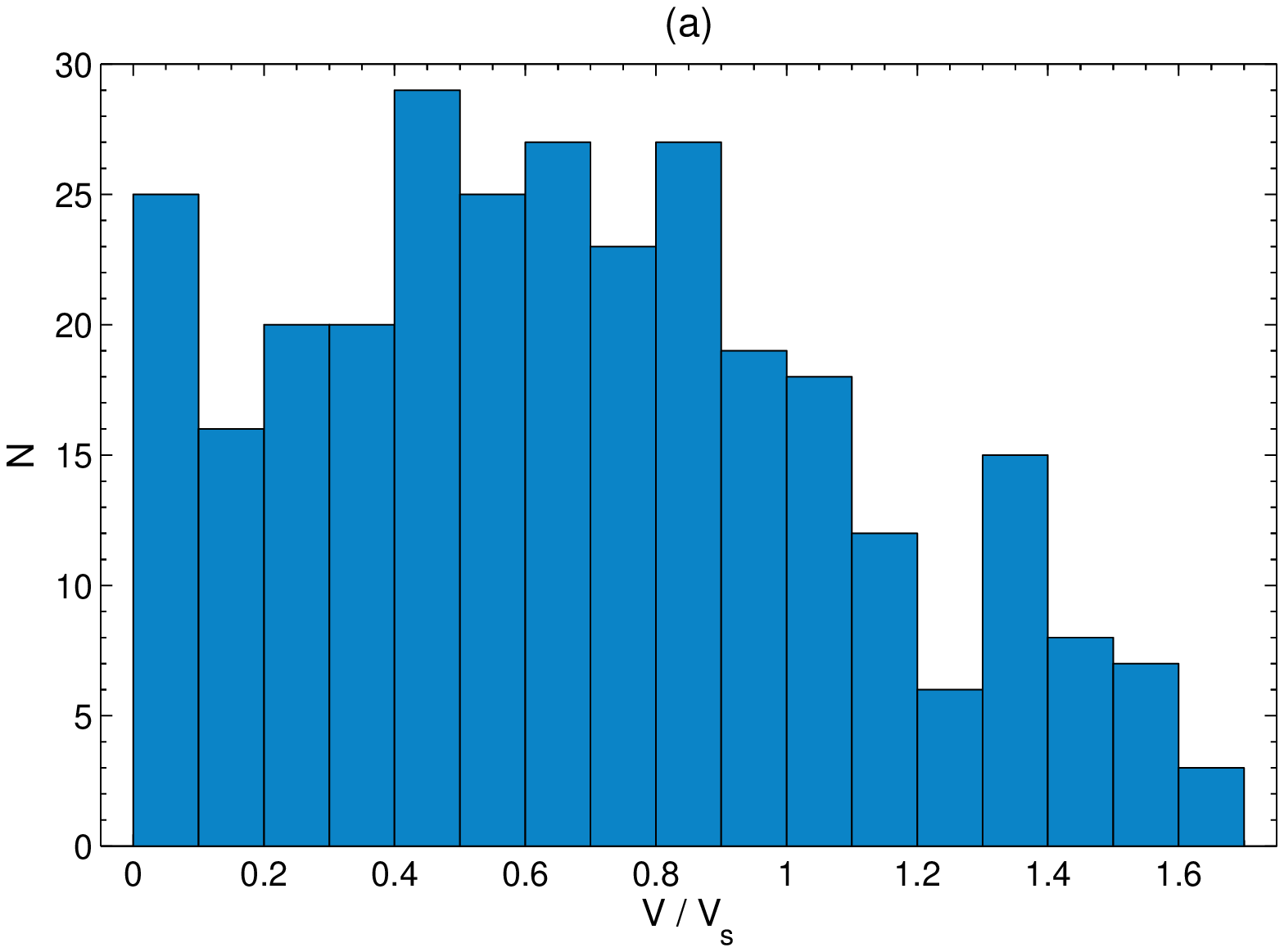}
\hspace{1cm}
\includegraphics[width=6cm]{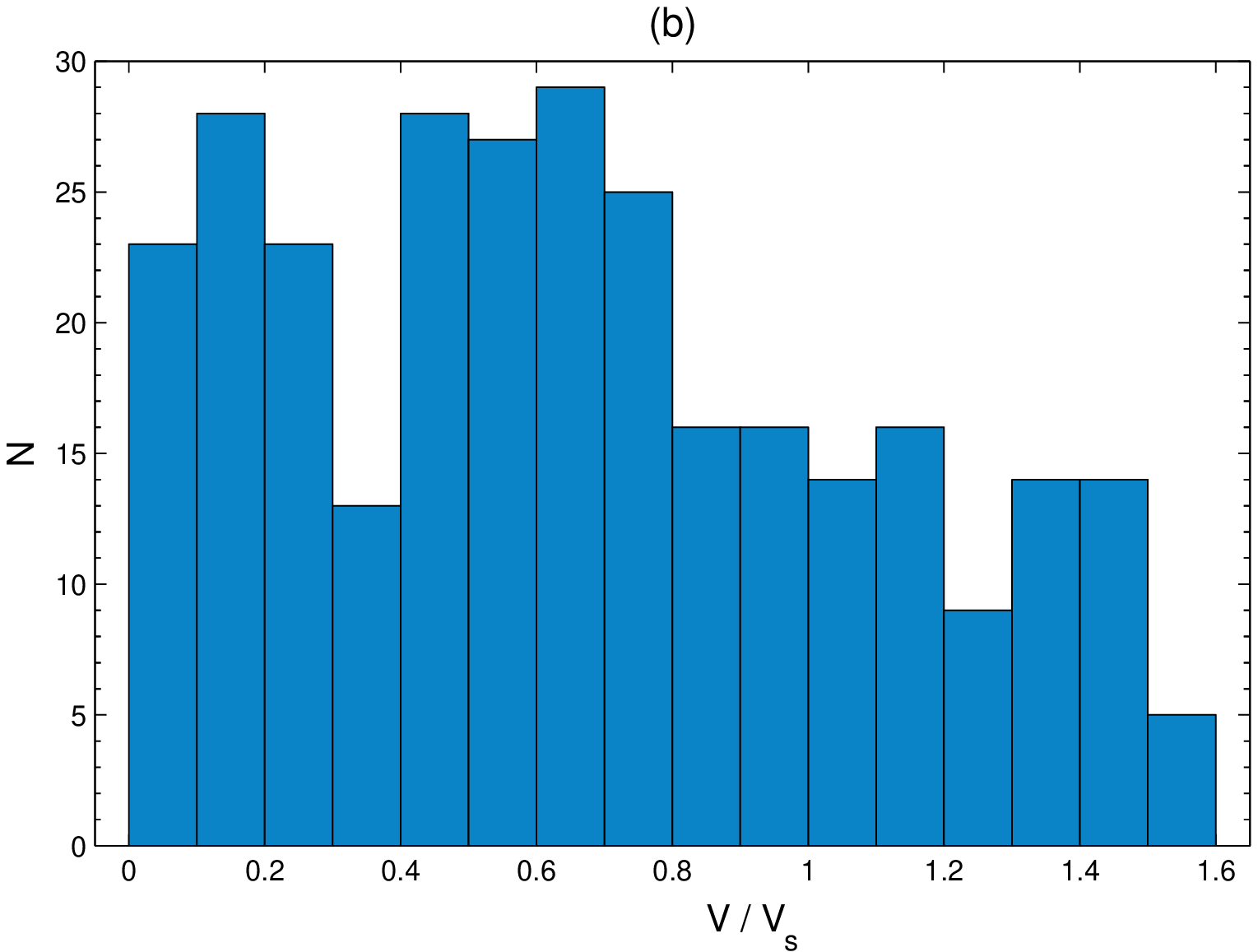}
\end{minipage}
\end{center}
\caption{Statistic of the visibility $V$ for the initial radical
pair state randomly chosen among the general product states (a) and
the incoherent states (b), compared with the visibility for the
singlet state $V_{s}$.} \label{Vis}
\end{figure}

\textsf{Magnetic field sensitivity from random separable or
incoherent states.---} Here we compare the magnetic field
sensitivity obtained from an initial singlet state (entangled) with
the sensitivity obtainable from classically correlated states
(separable). A mixed quantum state is called separable if it can be
written as a sum of product states
\begin{equation}
\rho=\sum\limits_{k}p_{k}|\phi_{k}\rangle_{a}\langle
\phi_{k}|\otimes|\psi_{k}\rangle_{b}\langle \psi_{k}|
\end{equation}
In general, a separable state can exhibit ``coherence'', by which
one means that some of its off-diagonal density matrix elements
(with respect to the standard basis
$\left\vert\uparrow\right\rangle$,
$\left\vert\downarrow\right\rangle$) are non-zero. By definition,
however, a separable state is not entangled. So there is an
essential difference between entanglement and coherence.

We introduce the optimal magnetic field sensitivity for the radical
pair reaction [Py-$h_{10}^{\cdot -}$ DMA-$h_{11}^{\cdot +}$] on the
set of separable states as
\begin{equation}
\Lambda_{Sep}(B)=\max\limits_{\rho\in Sep} |\Lambda(\rho,B)|
\end{equation}
where $\Lambda(\rho,B)$ denotes the magnetic-field sensitivity for a
given initial state $\rho$. It can be proved that the optimal
sensitivity $\Lambda_{Sep}(B)$ is obtained by the product states
$|\phi\rangle_{a}\otimes |\psi\rangle_{b}$. In our numerical
calculations, we randomly choose $5000$ product states and calculate
$\Lambda_{Sep}(B)$. We also randomly choose the initial state from
the set of incoherent states, the off-diagonal matrix elements of
which are all zero, meaning that no coherence is present (and
naturally no entanglement either). This is what we have done to
compute the curve `Sep' in Fig.~L1(b). We find, somehow surprising,
that the optimal sensitivity from the incoherent states
$\Lambda_{Inc}$ is the same as $\Lambda_{Sep}$. In this sense,
quantum coherence is not vital for the magnetic-field sensitivity.

In the example of FADH$^\bullet$-O$_{2}^{\bullet -}$, we
characterize the angle dependence by using the quantity of
visibility defined as follows
\begin{equation}
V=\frac{\max \Phi_{s}-\min \Phi_{s}}{\max \Phi_{s}+\min \Phi_{s}}
\end{equation}
where $\Phi_{s}$ is the singlet yield. We have randomly chosen a few
hundreds of product states and incoherent states as the initial
radical pair state, and calculated the corresponding visibility $V$
(compared with the visibility $V_{s}$ for the singlet state). It can
be seen from Fig.~\ref{Vis} that a substantial part of separable
(incoherent) states can account for an angular dependence that is as
high as (or even higher) than for the singlet state.  In this sense,
the radical pair initial state of an avian compass need not be the
singlet state.\newline

\textsf{Quantum entanglement and magnetic field sensitivity.---} To
quantify the amount of entanglement that exists in the active
radical pairs during the reaction, similar to the singlet yield, we
define the effective entanglement $\Phi _{E}$ as the integral
\begin{equation}
\Phi _{E}=\int_{0}^{\infty }r_{c}(t)E(t)dt
\end{equation}
Its first derivative with respect to the magnetic field is the
entanglement sensitivity
\begin{equation}
\Lambda _{E}=\frac{\partial \Phi _{E}}{\partial B}
\end{equation}

\begin{figure}[tbh]
\epsfig{file=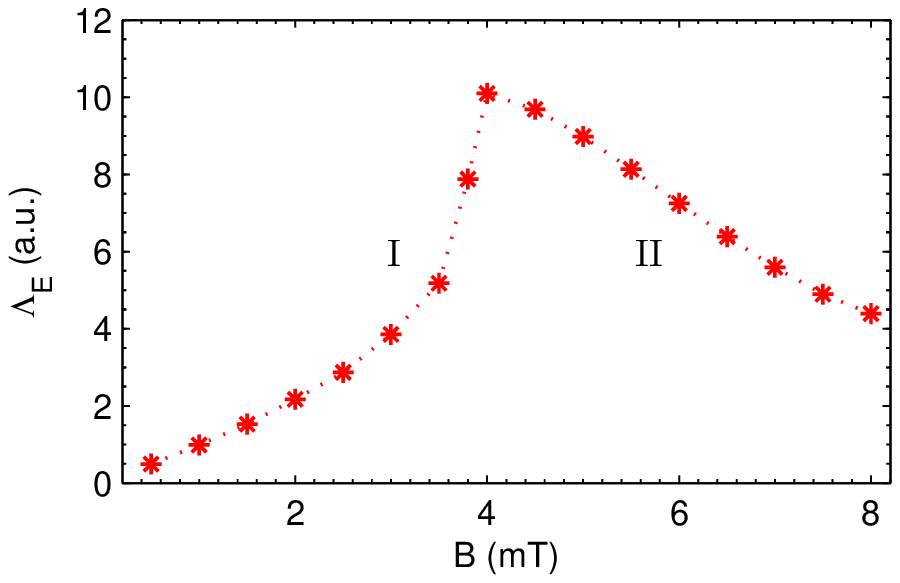,width=8cm} \caption{(Color online) Entanglement
sensitivity $\Lambda _{E}$ of a radical pair reaction
[Py-$h_{10}^{\cdot -}$ DMA-$h_{11}^{\cdot +}$] as a function of the
magnetic field $B$. The recombination rate constant is $k=5.8\times
10^{8}s^{-1}$ \protect\cite{Hor07s}.} \label{dEaE}
\end{figure}

In Fig.~\ref{dEaE}, we plot $\Lambda _{E}$ as a function of $B$. It
can be seen that $\Lambda _{E}$ changes conspicuously (kink) during
the crossover between the regions of I and II. At the same time, the
entanglement yield always increases with the magnetic field. This
can be understood from the fact that strong magnetic fields will
energetically suppress the relaxation (spin flips) in the
longitudinal direction. By this process, the state of the electron
pairs changes towards a binary mixture of two entangled states
(namely $|\mathbb{S}\rangle$ and $|\mathbb{T}_{0}\rangle$), which is
entangled for almost all values of the mixing parameter, resulting
in a much longer lifetime of entanglement.

\begin{figure}[tbh]
\begin{center}
\begin{minipage}{15cm}
\hspace{-0.5cm}
\includegraphics[width=7cm]{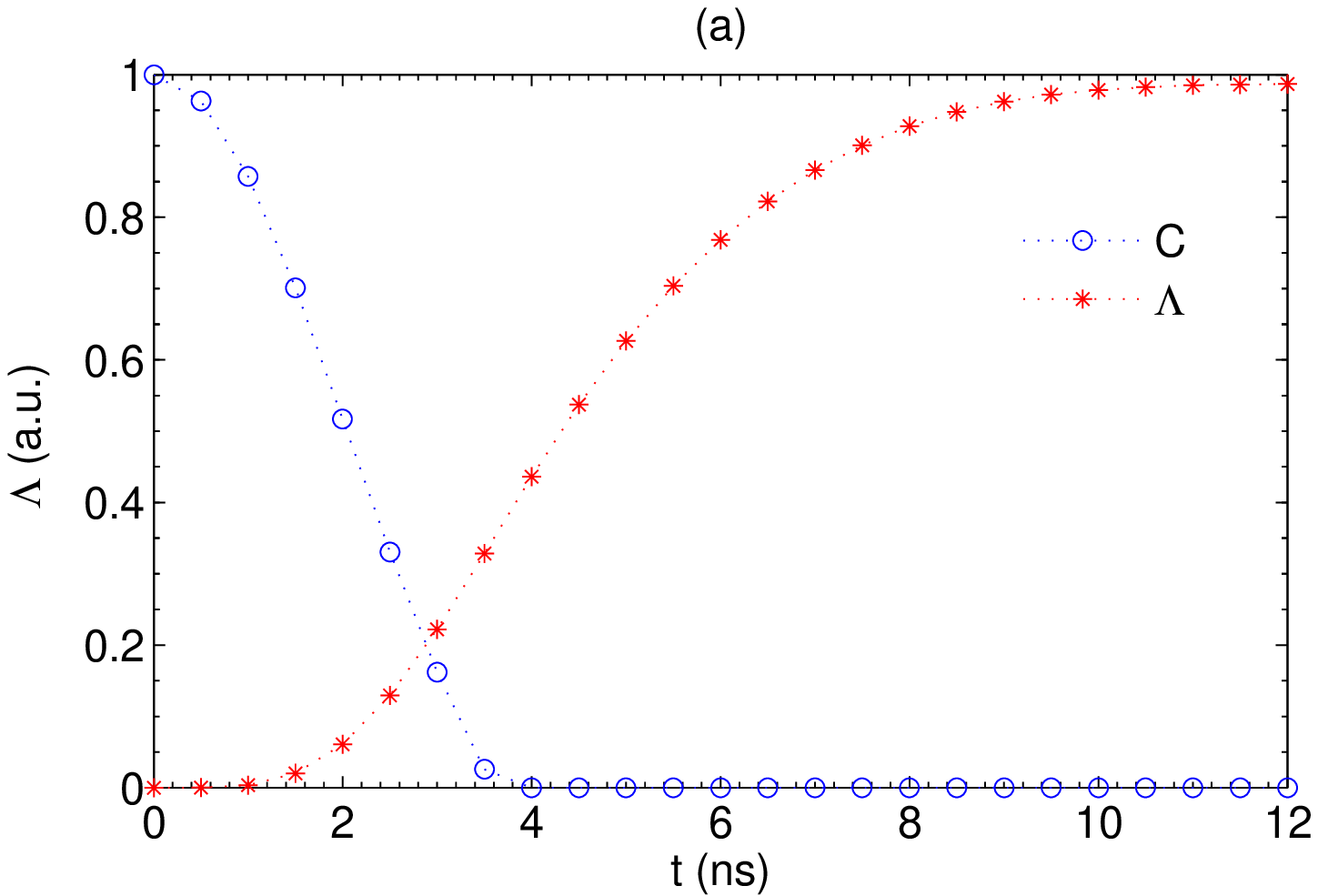}
\hspace{0.2cm}
\includegraphics[width=7cm]{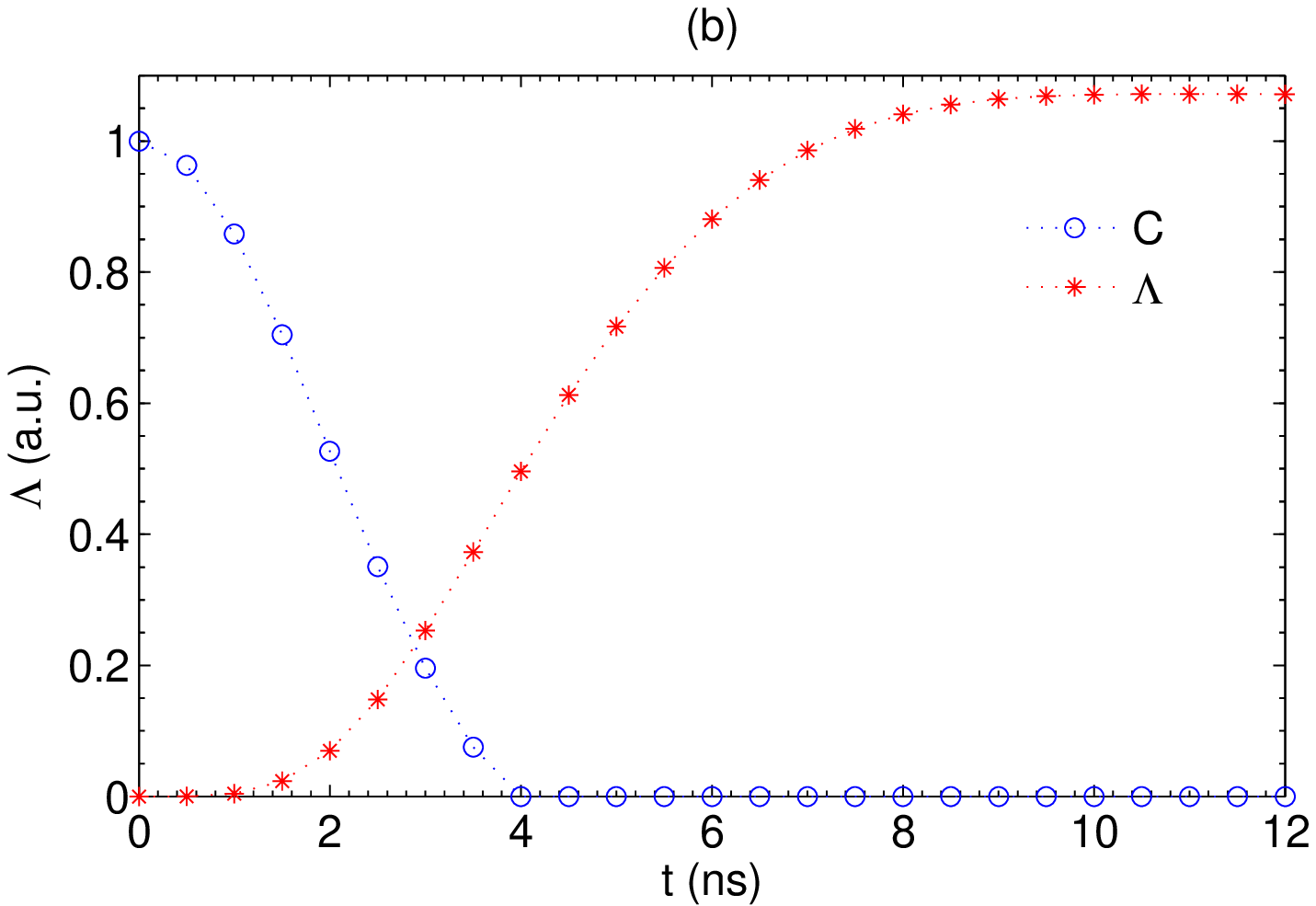}
\end{minipage}
\begin{minipage}{15cm}
\hspace{-0.5cm}
\includegraphics[width=7cm]{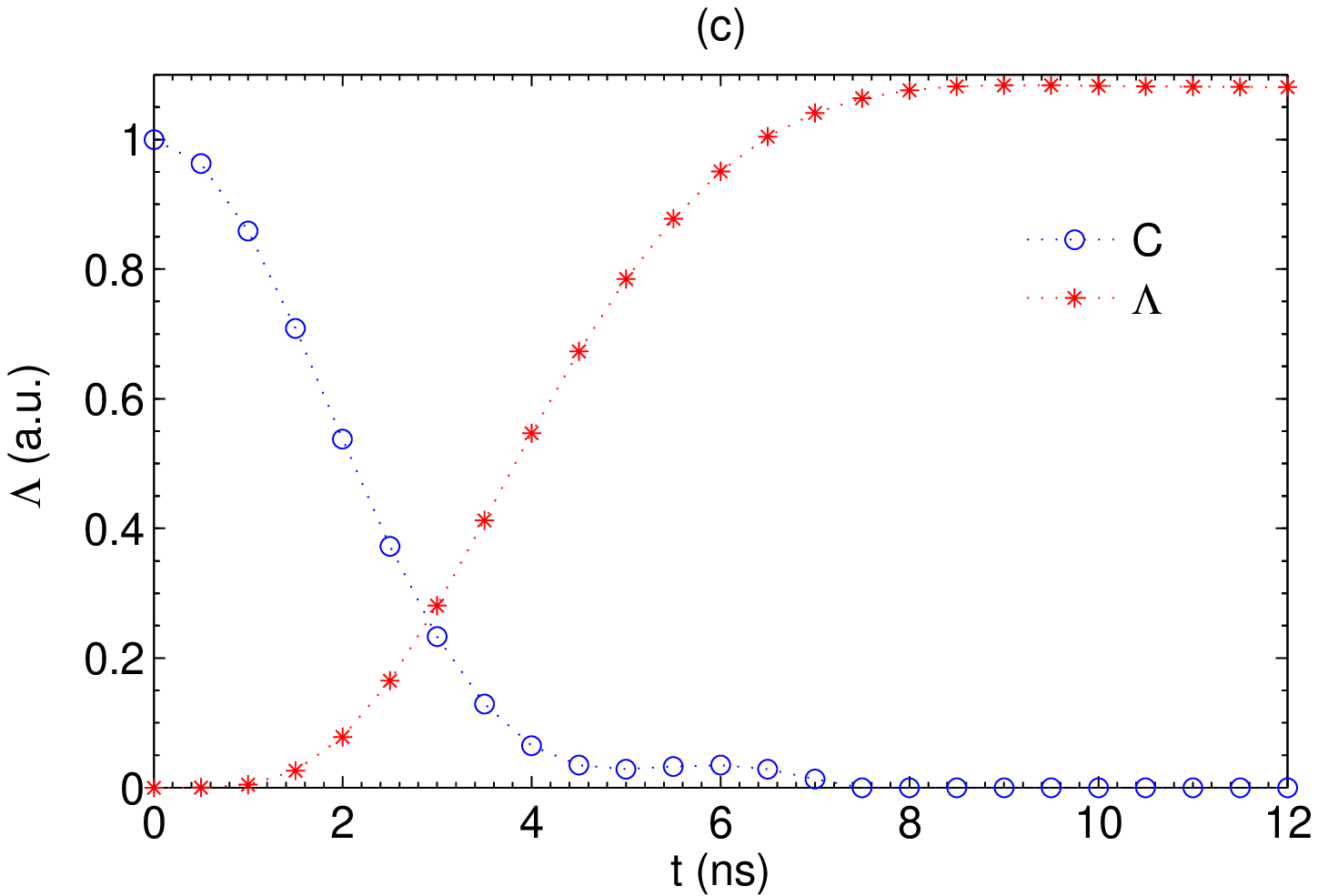}
\hspace{0.2cm}
\includegraphics[width=7cm]{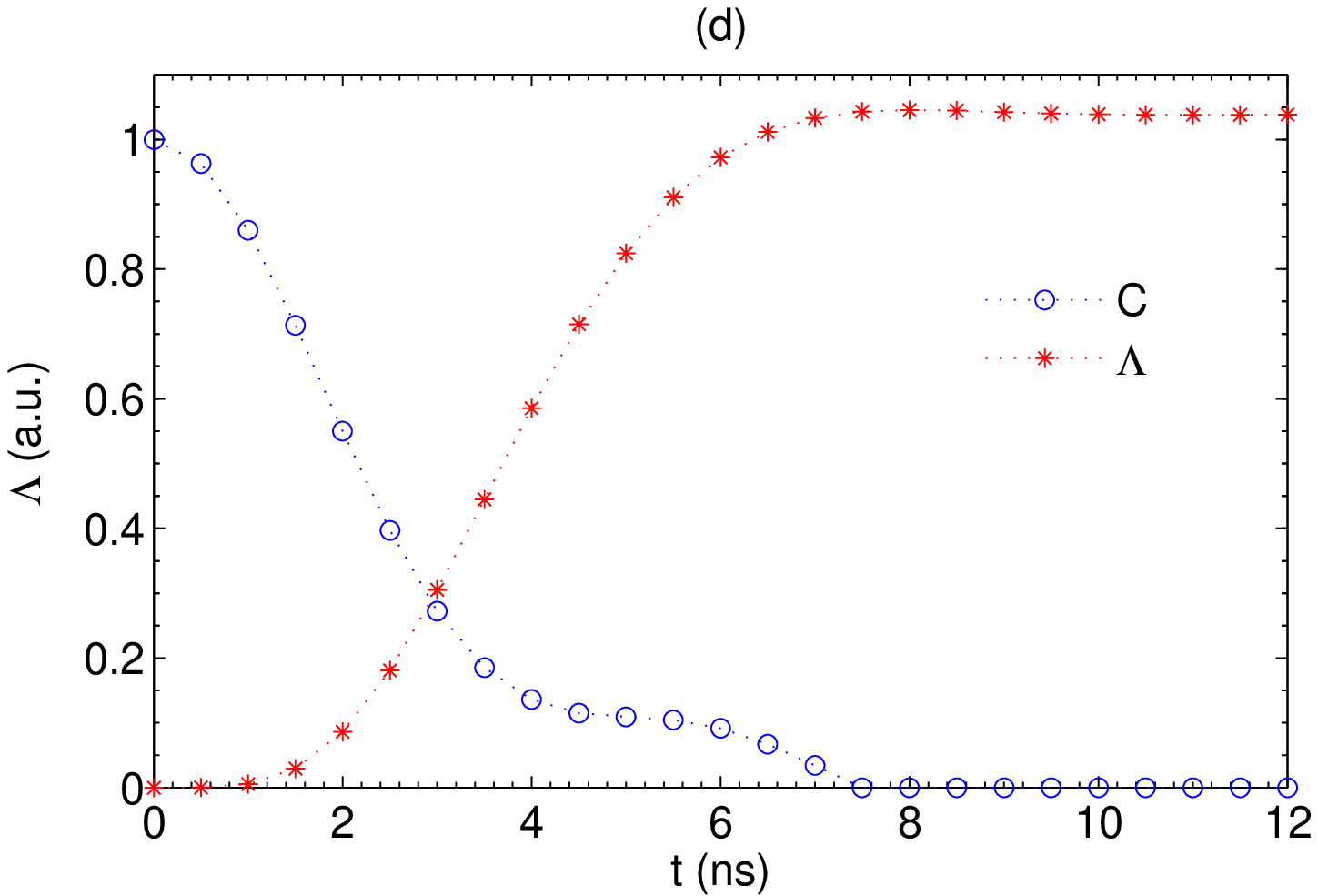}
\end{minipage}
\end{center}
\caption{(Color online) Time evolution of the (accumulated)
magnetic-field sensitivity $\Lambda(B,t)$ (rescaled) and the
entanglement of the radical pair for (a) $B=3$mT, (b) $3.5$mT, (c)
$4$mT, and (d) $4.5$mT in the radical pair reaction
[Py-$h_{10}^{\cdot -} $DMA-$h_{11}^{\cdot +}$]. The recombination
rate constant is $k=5.8\times 10^8$ s$^{-1}$ \protect\cite{Hor07s}.
} \label{pe}
\end{figure}

To further illustrate the connection between quantum entanglement
and the magnetic field sensitivity, we plot in Fig.~\ref{dEaE} the
time evolution of the entanglement and of the value of the
accumulated magnetic-field sensitivity $\Lambda(B,t)=\partial
\Phi(t)/\partial B$, for different values of the magnetic field:
$B=3$mT, $3.5$mT, $4$mT, and $4.5$mT. The lifetime of entanglement
in the region of I is approximately $\mathrm{T}_{E}=4$ns, while
$\Lambda(B,t)$ needs about $\mathrm{T}_{r}=10$ns to reach its
saturate value, see Fig.~\ref{pe}(a-b). We can also explicitly see
the sudden increase of $\mathrm{T}_{c}$ when $B$ crosses between the
regions I (low magnetic field) and II (high magnetic field), from
$\mathrm{T}_{E}=4$ns to about $7.3$ns, see Fig.~\ref{pe}(c-d), which
gives rise to the steps in Fig.~L2 (b) \cite{Note1}.

\emph{At this point, it is worth to emphasize that the concept of
entanglement is different from the singlet fraction (fidelity),
which was studied earlier. Generally speaking, a state can exhibit
significant classical spin correlations without having any
entanglement. In our specific example, at any time during the
radical-pair reaction there will be a finite singlet fraction while
entanglement, in contrast, will only exist for a much shorter time
(as shown in Fig.~\ref{pe}). In this sense, entanglement is a
different, and generally more sensitive, signature than the singlet
fraction. See also Fig.~L2(b) in the main text.}\newline

\textsf{Reference model of a bosonic heat bath.---} Let us assume
that each of the unpaired electron spins is coupled with an
independent bosonic heat bath at the same temperature. The dynamics
of one central spin would thus be
described by the following Lindblad type master equation \cite%
{BreuerBooks,Hans93s}
\begin{equation}
\frac{\partial }{\partial t}\rho =-i[H_{c},\rho
]+\sum_{k}(2L_{k}\rho L_{k}^{\dag }-\rho L_{k}^{\dag
}L_{k}-L_{k}^{\dag }L_{k}\rho )
\end{equation}%
where $H_{c}=m_{b}S_{z}$, with $m_{b}=-\gamma _{e}B$, $L_{1}=\sqrt{\gamma s}%
\sigma _{+}$ and $L_{2}=\sqrt{\gamma (1-s)}\sigma _{-}$. The
solution of the
above master equation can be represented by a map $\rho (t)=\overline{%
\mathcal{M}}_{t}[\rho (0)]$ which is explicitly expressed as follows
\begin{eqnarray*}
\overline{\mathcal{M}}_{t}:\quad \left\vert \uparrow \right\rangle
\left\langle \uparrow \right\vert &\rightarrow &\alpha
_{t}\left\vert \uparrow \right\rangle \left\langle \uparrow
\right\vert +(1-\alpha _{t})\left\vert \downarrow \right\rangle
\left\langle \downarrow \right\vert
\\
\left\vert \downarrow \right\rangle \left\langle \downarrow
\right\vert &\rightarrow &(1-\beta _{t})\left\vert \uparrow
\right\rangle \left\langle \uparrow \right\vert +\beta
_{t}\left\vert \downarrow \right\rangle
\left\langle \downarrow \right\vert \\
\left\vert \uparrow \right\rangle \left\langle \downarrow
\right\vert &\rightarrow &e^{-i2m_{b}t}\text{ }\eta _{t}\left\vert
\uparrow
\right\rangle \left\langle \downarrow \right\vert \\
\left\vert \downarrow \right\rangle \left\langle \uparrow
\right\vert &\rightarrow &e^{i2m_{b}t}\text{ }\eta _{t}\left\vert
\downarrow \right\rangle \left\langle \uparrow \right\vert
\end{eqnarray*}%
where $\alpha _{t}=(1-s)e^{-2\gamma t}+s$, $\beta _{t}=se^{-2\gamma
t}+(1-s)$ and $\eta _{t}=e^{-\gamma t}$. This map describes
spin-exchange interactions with the environment with an effective
rate $\gamma $ and an equilibrium parameter $s$ that is related to
the environment temperature $T$. The dependence of $\gamma $ and $s$
on $T$ and the magnetic field $B$ is given
in the following way: $\gamma =2m_{b}\kappa _{0}(2\mathcal{N}+1)$ and $s=%
\mathcal{N}/(2\mathcal{N}+1)$, where $\kappa _{0}$ depends on the
system-bath coupling strength on resonance, and the bosonic
distribution function is $\mathcal{N} =1/(e^{\frac{\epsilon
_{s}}{\epsilon _{T}}}-1)$ with the system energy scale $\epsilon
_{s}=2\hbar m_{b}$ and the thermal energy scale $\epsilon
_{T}=k_{b}T$. Thus we have
\begin{eqnarray}
\frac{1}{s}\frac{\partial s}{\partial B} &=&-\frac{s}{B}\frac{\epsilon _{s}}{%
\epsilon _{T}}e^{\frac{\epsilon _{s}}{\epsilon _{T}}} \\
\frac{1}{\gamma }\frac{\partial \gamma }{\partial B} &=&\frac{1}{B}\left[ 1-2%
\frac{\epsilon _{s}}{\epsilon _{T}}e^{\frac{\epsilon _{s}}{\epsilon _{T}}%
}(e^{2\frac{\epsilon _{s}}{\epsilon _{T}}}-1)^{-1}\right]
\end{eqnarray}

We are interested in the effects of low magnetic fields, for example
$B=1$ mT, which corresponds to the thermal energy scale at
temperature $T\simeq 2.69$ mK that is quite low for biochemical
systems. Thus we can naturally assume that $\frac{\epsilon
_{s}}{\epsilon _{T}}\ll 1$, from which it is
easy to verify that $\left\vert \frac{1}{\gamma }\frac{\partial \gamma }{%
\partial B}\right\vert \ll \left\vert \frac{1}{s}\frac{\partial s}{\partial B%
}\right\vert $, e.g. if $T=1$K then $\left\vert \frac{1}{\gamma }\frac{%
\partial \gamma }{\partial B}\right\vert $ is already four orders smaller
than $\left\vert \frac{1}{s}\frac{\partial s}{\partial B}\right\vert
$.

\begin{figure}[tbh]
\epsfig{file=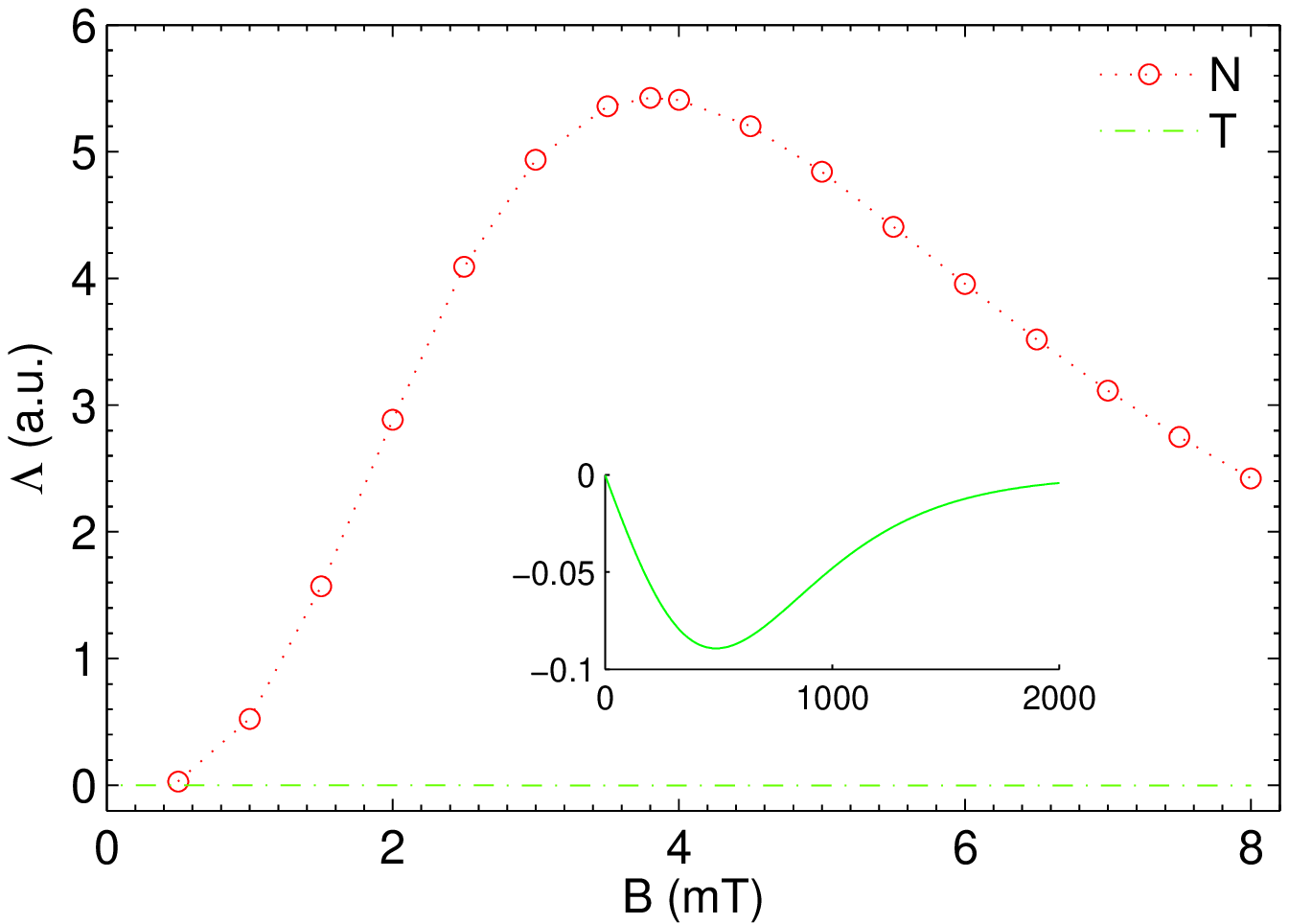,width=8cm} \caption{(Color online)
Magnetic field sensitivity $\Lambda$ resulting from
the nuclear spin environment ($N$) of the radical pair reaction [Py-$%
h_{10}^{\cdot -}$ DMA-$h_{11}^{\cdot +}$]; and the optimal $\Lambda$
(achieve when $\protect\gamma \gg k$) from the bosonic heat bath at
temperature $T=1$K (see also Inset for an extended range of
parameter) as a
function of the magnetic field $B$. The recombination rate constant is $%
k=5.8\times 10^{8}$s$^{-1}$ \protect\cite{Hor07s}. } \label{ThB}
\end{figure}

The radical pair starts in the singlet state $|\mathbb{S}\rangle =\frac{1}{%
\sqrt{2}}(\left\vert \uparrow \downarrow \right\rangle -\left\vert
\downarrow \uparrow \right\rangle )$, and its state evolves as $\rho _{s}(t)=%
\overline{\mathcal{M}}_{t}^{(1)}\otimes \overline{\mathcal{M}}_{t}^{(2)}[%
\mathcal{P}_{s}]$. At time $t$, the density matrix is of the
following form
\begin{equation}
\rho _{s}(t)=\left(
\begin{array}{cccc}
a & 0 & 0 & 0 \\
0 & b & c & 0 \\
0 & c & b & 0 \\
0 & 0 & 0 & d%
\end{array}%
\right)  \label{STB}
\end{equation}%
where $a=\alpha _{t}(1-\beta _{t})$, $b=\left[ \alpha _{t}\beta
_{t}+(1-\alpha _{t})(1-\beta _{t})\right] /2$, $d=(1-\alpha _{t})\beta _{t}$%
, and $c=-\eta _{t}^{2}/2$. Thus we can calculate the singlet fidelity $%
f_{s}(t)=\mathrm{Tr}\left[ \rho (t)\mathcal{P}_{s}\right] =b-c$ as
\begin{equation*}
f_{s}(t)=\frac{1}{2}\left[ \alpha _{t}\beta _{t}+(1-\alpha
_{t})(1-\beta _{t})+\eta _{t}^{2}\right]
\end{equation*}%
The activation yield for the exponential re-encounter probability model is $%
\Phi =\int_{0}^{\infty }f_{s}(t)ke^{-kt}dt$, i.e.
\begin{equation*}
\Phi =\frac{k}{k+2\gamma }+\frac{8\gamma ^{2}}{(k+4\gamma )(k+2\gamma )}%
s(1-s)
\end{equation*}%
One can verify that under the general conditions we are interested
in, the magnitude of the magnetic field sensitivity $\Lambda $ would
increase with the coupling strength scale $\kappa _{0}$, i.e. the
fast thermalization is good in the present context. To achieve the
optimal bound of $\Lambda $ and illustrate the essential physics, we
can assume that $\gamma $ is much larger than $k$ (this is different
from the real situation where $\gamma$ is smaller than $k$), which
leads to
\begin{equation}
\Lambda \simeq -\left( 1-2s\right) \frac{s^{2}}{B}\frac{\epsilon _{s}}{%
\epsilon _{T}}e^{\frac{\epsilon _{s}}{\epsilon _{T}}}
\end{equation}%
The magnitude of $\Lambda $ from the bosonic heat bath decreases as
the temperature increases. Even at temperature as low as $1$ K, it
is already significantly smaller than the one from the nuclear spin
environment, see Fig.~\ref{ThB}. Therefore, we can conclude that the
effects of low magnetic fields will indeed be washed out completely
by the thermal fluctuations.

By calculating $\partial |\Lambda |/\partial B$, we find that
$\left\vert \Lambda \right\vert $ will always grow as the magnetic
field becomes stronger, as long as $\frac{\epsilon _{s}}{\epsilon
_{T}}\leq \ln (2+\sqrt{3} )$, which is obviously satisfied in the
regions we are interested in. The change of the sign of $\partial
|\Lambda |/\partial B$ happens at $\frac{\epsilon _{s}}{\epsilon
_{T}}=\ln (2+\sqrt{3})$. At room temperature $T=300$ K, this would
correspond to the magnetic field $B\sim 135$T.

\begin{figure}[tbh]
\epsfig{file=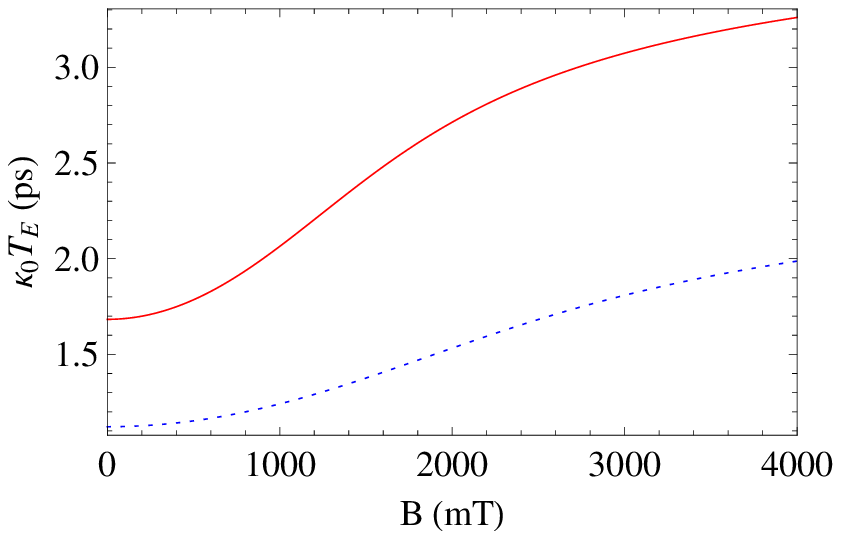,width=8cm} \caption{(Color online)
Lifetime of entanglement $\protect\kappa_{0} T_{E}$ as a function of
the magnetic field $B$. The bosonic thermal bath temperature is
$T=1$ K (red solid) and $T=1.5$ K (blue dotted).} \label{ELThB}
\end{figure}

The evolution of entanglement as obtained from Eq.~(\ref{STB}) is $%
E(t)=\max\{0,2(|c|-(ad)^{1/2})\}$. In a similar way, one can obtain
the lifetime of entanglement, see Fig.~\ref{ELThB}, which is
monotonically increasing with the magnetic field. This is another
feature in marked contrast with the nuclear spin environment: there
are no oscillatory kinks in the entanglement lifetime as the
magnetic field increases.

\end{document}